\documentclass[
 prl,
 nobibnotes,
 reprint,
 amsmath,amssymb,
 aps,
 showkeys,
 nofootinbib,
 superscriptaddress
]{revtex4-2}

\usepackage{graphicx}% Include figure files
\usepackage{dcolumn}% Align table columns on decimal point
\usepackage{bm}% bold math
\usepackage{ragged2e}
\usepackage{subcaption}   % “sub‑figure” support (works with RevTeX)
\usepackage{overpic}
\newcommand{\panel}[3][90,59]{%
  \begin{minipage}[t]{\plotwidth}\centering
    \begin{overpic}[width=\linewidth]{figures/#2.pdf}
      \put(#1){\textbf{(#3)}}
    \end{overpic}
  \end{minipage}}
\usepackage{multirow} % For multirow cells in tables
\usepackage{orcidlink}
\usepackage{xspace}

\newcommand{\dd}{\mathrm{d}}
\newcommand{\plotwidth}{0.37\linewidth}
\newcommand{\EPOSLHCR}{\texttt{EPOS-LHC-R}\xspace}
\newcommand{\SibyllTwoThreeE}{\texttt{Sibyll\,2.3e}\xspace}
\definecolor{rossoCP3}{cmyk}{0,.88,.77,.40}
\definecolor{darkBlue}{rgb}{0, 0, 0.8}
\hypersetup{
  bookmarks=true,         % show bookmarks bar?
  bookmarksopen=true,     % open bookmarks
  bookmarksopenlevel=1,
  unicode=false,          % non-Latin characters in Acrobat’s bookmarks
  pdftoolbar=true,        % show Acrobat’s toolbar?
  pdfmenubar=true,        % show Acrobat’s menu?
  colorlinks=true,        % false: boxed links; true: colored links
  linkcolor=darkBlue,     % color of internal links
  citecolor=darkBlue,     % color of links to bibliography
  filecolor=darkBlue,     % color of file links
  urlcolor=darkBlue       % color of external links
}
\begin{document}

% ----------------------------------------------------------------------
% TITLE
% ----------------------------------------------------------------------
\title{Ultra-high-energy cosmic rays require metal-rich sources}

% ----------------------------------------------------------------------
% AUTHORS & AFFILIATIONS
% ----------------------------------------------------------------------
\author{E.~Podlesnyi$^{\orcidlink{0000-0003-3395-0419}}$}\email{egor.podlesnyi@ntnu.no}
\affiliation{
 Department of Physics, Norwegian University of Science and Technology (NTNU),\\
 Høgskoleringen 5, Trondheim 7491, Norway
}

\author{F.~Oikonomou$^{\orcidlink{0000-0002-0525-3758}}$}
\affiliation{
 Department of Physics, Norwegian University of Science and Technology (NTNU),\\
 Høgskoleringen 5, Trondheim 7491, Norway
}

\author{D.~Ehlert$^{\orcidlink{0000-0002-4322-6400}}$}
\affiliation{
 Department of Physics, Norwegian University of Science and Technology (NTNU),\\
 Høgskoleringen 5, Trondheim 7491, Norway
}

\author{M.~Unger$^{\orcidlink{0000-0002-7651-0272}}$}
\affiliation{
 Institut f\"ur Astroteilchenphysik, Karlsruher Institut für Technologie,\\
 Karlsruhe D‑76344, Germany
}
\affiliation{
 Department of Physics, Norwegian University of Science and Technology (NTNU),\\
 Høgskoleringen 5, Trondheim 7491, Norway
}

% ----------------------------------------------------------------------
% DATE
% ----------------------------------------------------------------------
\date{\today}

% ----------------------------------------------------------------------
% ABSTRACT
% ----------------------------------------------------------------------
\begin{abstract}
The elemental composition of ultra-high-energy cosmic rays (UHECRs) must be evaluated at a given rigidity rather than energy to compare with candidate elemental abundances, such as those of low-energy cosmic rays seeding the acceleration of UHECRs. We determine the UHECR composition from fits to Pierre Auger Observatory data and evaluate the resulting abundances at fixed rigidity. We identify two classes of solutions consistent with measurements above $10^{17.8}\,\mathrm{eV}$: one in which the hydrogen flux at Earth is dominated by secondary nucleons produced by photodisintegration of heavier nuclei in extragalactic photon fields, and a second in which this contribution is less pronounced. Only the latter, which yields a mixture of primary and secondary protons at Earth, allows a hydrogen abundance at the source comparable to that of low-energy cosmic rays. However, both classes of solutions require an overabundance of nuclei in the silicon-iron group. Other source models from the literature also fail to reproduce the full elemental composition. The metallicity of UHECRs thus differs markedly from that of all candidate seed populations considered, leaving the origin of the heavy-element excess unresolved.
\end{abstract}
\maketitle

% ----------------------------------------------------------------------
% MAIN TEXT
% ----------------------------------------------------------------------
\textit{Introduction}---The elemental composition of ultra-high-energy cosmic rays (UHECRs), atomic nuclei with energies $E \gtrsim 10^{18}$~eV, provides important clues to their still-unknown sources. It reflects both the material available for acceleration and the conditions in the production site.

Candidate sources include radio galaxies, gamma-ray bursts, tidal disruption events, galaxy clusters, and binary neutron star mergers (for reviews, see, e.g., \cite{Bister:2025tse, Globus:2025ftu,AlvesBatista:2019tlv, Anchordoqui:2018qom}). These environments provide different seed material for acceleration, ranging from pre-existing cosmic rays to stellar ejecta. One possibility is that UHECRs are produced by reaccelerating lower-energy cosmic rays (CRs) present in the host environments, similar to Galactic Cosmic Rays (GCRs) observed in the Milky Way~\cite{Caprioli:2015zka, Kimura:2017ubz, Simeon:2025gxd}.

Interpreting measurements of the UHECR composition at Earth requires understanding the elemental abundances expected at the sources, accounting for interactions during propagation. A comparison with candidate seed populations must be made at fixed rigidity $R = E/Z$ rather than fixed energy $E$, where $Z$ is the CR charge. This is particularly important for populations with different power-law spectral indices, such as GCRs with soft spectra and UHECRs with hard ones, since, as we demonstrate in this Letter, the common comparison at fixed energy can substantially underestimate the UHECR enrichment in heavy nuclei.

We infer the elemental abundances of UHECRs escaping their sources from fits to the UHECR spectrum~\cite{PierreAuger:2025eun} and composition measurements~\cite{PierreAuger:2026qbt} obtained by the Pierre Auger Observatory~\cite{PierreAuger:2015eyc} (hereafter, \textit{Auger}). We find that UHECR sources must be strongly enriched in nuclei of the silicon--iron group relative to GCRs, while their hydrogen abundances can be compatible. This excess persists independent of experimental and hadronic-interaction uncertainties and of source evolutions considered. Our results disfavour simple rigidity-dependent reacceleration of GCR-like seed populations and place stringent constraints on the composition required of candidate UHECR sources.

\textit{Elemental fractions at a given rigidity}---The energy spectrum of CR nuclei of type $i$ with charge $Z_{i}$ can be described as a power law with a superexponential cutoff of power $\lambda$:
\begin{equation}
    \left(\frac{dN}{dE} \right)_{i,\alpha} = C_{i,\alpha} \left( \frac{E}{E_{\mathrm{ref}}} \right)^{-\gamma_{\alpha}} \exp \left[- \left( \frac{E}{E^{\mathrm{cut}}_{i, \alpha}} \right)^{\lambda} \right].
\label{eq:source_spectrum}
\end{equation}
Here, $E_{\mathrm{ref}}$ is a constant reference energy, $C_{i, \alpha}$ is the normalisation of the CR spectrum at the reference energy, the index $\alpha$ is needed to distinguish between different power laws with different spectral indices $\gamma_{\alpha}$ ($\alpha = 1$ means ``before reacceleration'', and $\alpha=2$ after), and $E^{\mathrm{cut}}_{i, \alpha}$ is the cutoff energy. We define the relative abundance (fraction) of nucleus $i$ at a given \textit{energy} as $\varepsilon_{i,\alpha}~=~C_{i,\alpha}/\sum_{k} C_{k,\alpha}$.

In a magnetic field, the motion of a relativistic nucleus depends on $E$ and $Z$ only through its rigidity $R$, which sets the Larmor radius. Acceleration, confinement and escape governed by magnetic fields therefore act on all nuclei in the same way at the same rigidity, so that the cutoff energy scales as $E^{\mathrm{cut}} \propto Z$ (the so-called Peters cycle~\cite{Peters:1961mxb}). This applies to diffusive shock acceleration as well as to the reacceleration models proposed for UHECRs, such as ``Espresso'' \cite{Caprioli:2015zka} and discrete shear acceleration \cite{Rieger:2004jz, Rieger:2006uv, Rieger:2019hrf, Kimura:2017ubz}. In these models, only CRs whose Larmor radius exceeds the width of the boundary of the reaccelerating region can cross it, i.e. only CRs above a rigidity threshold are reaccelerated. The rigidity scaling is broken only by processes that depend on the mass number $A$ and the charge $Z$ separately, such as photodisintegration and other in-source interactions (discussed below).

Elemental abundances should therefore be compared at the same rigidity rather than the same energy. For spectra following Eq.~(\ref{eq:source_spectrum}), the fractions $\rho_i$ at a given rigidity are related to the fractions $\varepsilon_i$ at a given energy by
\begin{equation}
    \rho_{i} = \frac{\displaystyle{\varepsilon_{i} Z_{i}^{1 - \gamma}}}{\displaystyle{\sum \limits_{i} \varepsilon_{i}  Z_{i}^{1 - \gamma}}}.
\label{eq:fractions_energy_to_rigidity}
\end{equation}
Rigidity-dependent reacceleration therefore conserves the fractions $\rho_{i}$, but not the fractions $\varepsilon_{i}$. If all CRs above a rigidity threshold are reaccelerated, and the cutoff also scales with $R$, then $\rho_{i}$ is unchanged, while $\varepsilon_{i}$ changes whenever $\gamma_{1} \neq \gamma_{2}$\footnote{The proof for CRs with the spectrum following Eq.~(\ref{eq:source_spectrum}) is given in the Supplemental Material.}:
\begin{equation}
    \frac{\varepsilon_{i, 2}}{\varepsilon_{j, 2}} = \frac{\varepsilon_{i, 1}}{\varepsilon_{j, 1}} \left( \frac{Z_{i}}{Z_{j}} \right)^{\gamma_2 - \gamma_1}.
\label{eq:change_of_metallicity}
\end{equation}
\begin{figure}[t]
    \centering
    \includegraphics[width=\linewidth]{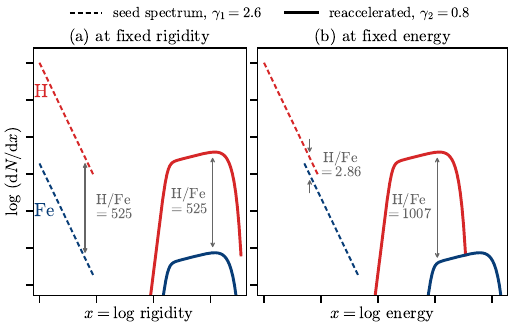}
\caption{
\justifying
 Schematic spectra of GCR-like H and Fe seeds ($\gamma_1 = 2.6$, thin dashed) and of the same nuclei after rigidity-dependent reacceleration ($\gamma_2 = 0.8$, thick solid), shown in rigidity (a) and in energy (b). Reacceleration leaves H/Fe at fixed rigidity unchanged, whereas at fixed energy it increases by a factor $Z_{\rm Fe}^{\gamma_1-\gamma_2} \approx 350$. \label{fig:schematic}}
\end{figure}
Thus, a reacceleration conserving the number of nuclei integrated from a given minimum \textit{rigidity} with a \textit{rigidity}-dependent cutoff conserves the fractions $\rho_{i}$ at a given \textit{rigidity} for any $\gamma_{1}$ and $\gamma_{2}$, while at a given \textit{energy} the boosted spectrum becomes \textit{harder} and \textit{lighter} (i.e. \textit{less metallic}) if $\gamma_{2} < \gamma_{1}$. This is the case relevant for UHECRs, as Galactic cosmic rays have a soft spectrum ($\gamma_{1} \approx 2.6$), whereas the spectra of UHECRs escaping their sources are predicted to be hard ($\gamma_{2} \approx 0.5$\text{--}$1$)~\cite{PierreAuger:2016use}. Reaccelerating GCR-like seeds would therefore conserve their $\rho_i$ while reducing the fractions $\varepsilon_i$ of heavy nuclei by $Z_i^{|\gamma_2-\gamma_1|}$ relative to light ones, so comparisons at a given energy underestimate any excess of heavy nuclei (see Fig.~\ref{fig:schematic}). Source abundances are, however, often quoted at a fixed energy (e.g.,~\cite{PierreAuger:2016use}) or as a fraction of the total energy carried by a given nucleus type above a fixed energy threshold (e.g.,~\cite{Plotko:2022urd}, see the Supplemental Material).

\textit{Data and methods}---To infer the metallicity of UHECRs, we use three data sets obtained by \textit{Auger}: (1)~the UHECR sky-averaged spectrum $J(E)$ measured with the surface detector of the Pierre Auger Observatory~\cite{PierreAuger:2025eun}; (2)~the mean values $\langle X_{\max} \rangle$ and standard deviations $\sigma \left( X_{\max} \right)$ of the extensive‑air‑shower maxima measured with the fluorescence detector of the Pierre Auger Observatory~\cite{PierreAuger:2026qbt}; (3) the H spectrum inferred from multiplying $J(E)$  from~\cite{PierreAuger:2021hun} with the reconstructed proton energy-dependent fractional abundances of~\cite{PierreAuger:2026qbt}. In this way, following \cite{Luce:2022awd}, we fit the full composition above the ankle, but avoid a potential contamination from GCRs at low energies and include only protons which are of guaranteed extragalactic origin due to anisotropy constraints~\cite{Giacinti:2011ww,Bister:2026kva}.

To simulate the intergalactic propagation of UHECRs, we use the same numerical framework as was utilised in Ref.~\cite{Ehlert:2022jmy} (see Sect. IIIB therein), which uses \texttt{CRPropa3}~\cite{CRPropa:2016qdt}.

The spectra of UHECRs escaping from the sources are assumed to follow a power-law with a \textit{rigidity}-dependent superexponential cutoff described by Eq.~(\ref{eq:source_spectrum}) with $E^{\mathrm{cut}}_{i} = Z_{i} R_{\mathrm{cut}}$, where the free parameter $R_{\mathrm{cut}}$ is the characteristic value of the rigidity cutoff, $i = $ H, He, N, Si, or Fe. The spectral index $\gamma$ (free parameter) and five fractions $\varepsilon_{i}$ at fixed \textit{energy} (four of which are free parameters) are used to infer fractions $\rho_{i}$ at fixed \textit{rigidity} following Eq.\:(\ref{eq:fractions_energy_to_rigidity}) (see the Supplemental Material for details regarding the fitting procedure).

The \texttt{CRPropa3} simulation provides the spectrum of UHECRs arriving on Earth with mass numbers $1{\leq}A{\leq}56$, with each $A$ traced individually. This allows us to compare the model prediction not only with the measured spectrum \cite{PierreAuger:2025eun} but also with the first two moments $\langle X_{\max} \rangle$ and $\sigma \left( X_{\max} \right)$ of the distribution of the extensive-air-shower maxima \cite{PierreAuger:2026qbt}. We use the recent hadronic interaction model (HIM) \EPOSLHCR \cite{Pierog:2023ahq} with the parametrisation given in Ref. \cite{Mollerach:2025fmx}. Details of our fitting procedure and (qualitatively similar) results for an alternative HIM \SibyllTwoThreeE \cite{Riehn:2019jet, Evoli:2026bea} are given in the Supplemental Material.

We explore various models for the UHECR cosmological evolution function $f(z, m)$. Following Ref.~\cite{vanVliet:2019nse}, we take $f(z,m) \propto (1+z)^{m}$ up to $z_{\max} = 4$, flattened above $z_{0} = 1.5$ for $m > 0$. We test values of $m~\in~\{-4; -2; 0; +2; +4; +5\}$. Besides, we test the phenomenological function of Madau \& Dickinson (see Eq.~(15) in Ref. \cite{Madau:2014bja}), which describes the star formation rate as a function of $z$ and, at lower redshifts, is similar to $f(z, 2.7)$.
\begin{figure}[t]
    \centering
    \includegraphics[width=0.9\linewidth]{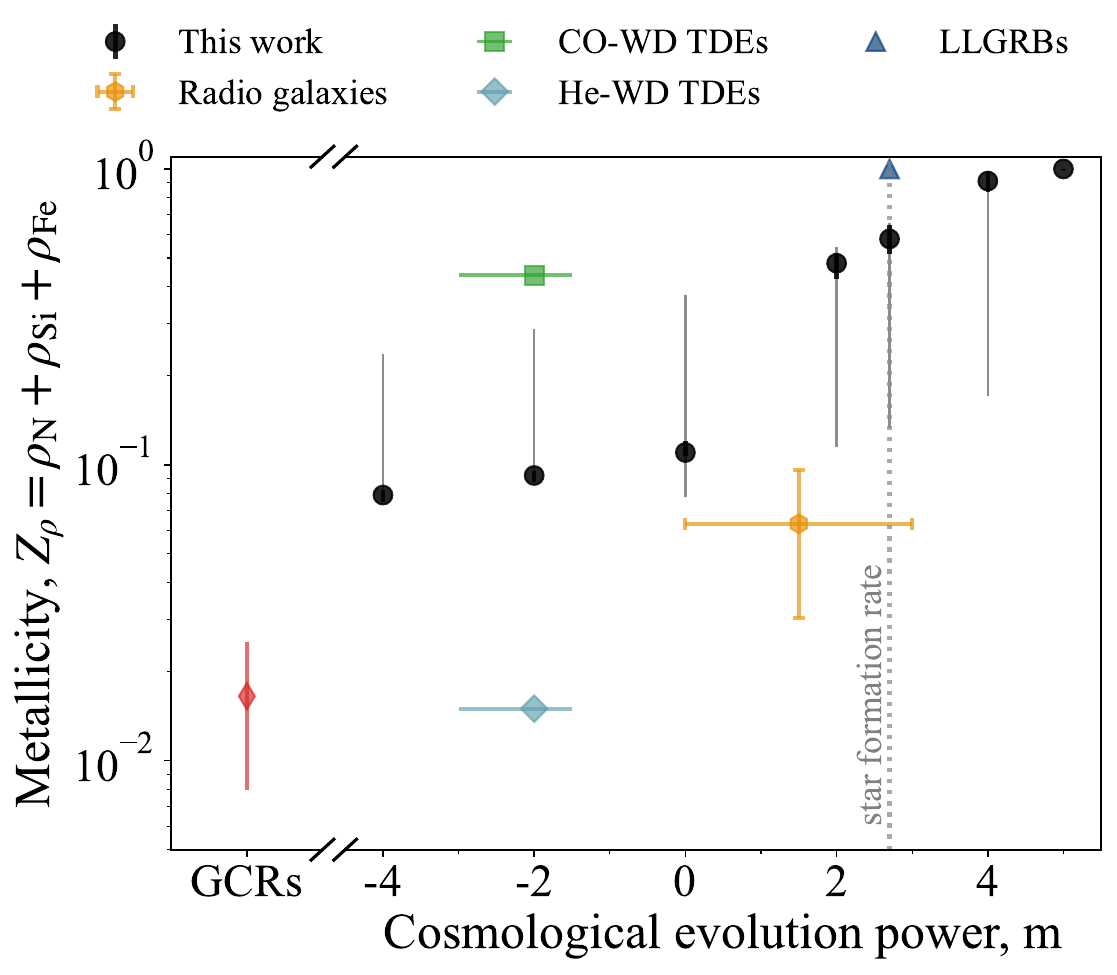}
    \caption{
    \justifying
        The metallicity of UHECRs $Z_{\rho}$ as a function of the power $m$ of the cosmological source evolution, in comparison with various models of UHECR sources. The metallicity of Galactic cosmic rays is also plotted as the red diamond. Black error bars show the 68\% highest-density credible intervals for the nominal \textit{Auger} data, while grey error bars show the spread of the credible intervals among the perturbations of the \textit{Auger} data, taking into account systematic uncertainties in $\langle X_{\max} \rangle$.
    }\label{fig:metallicity_as_function_of_m}
\end{figure}
\begin{figure}[t]
    \centering
    \includegraphics[width=\linewidth]{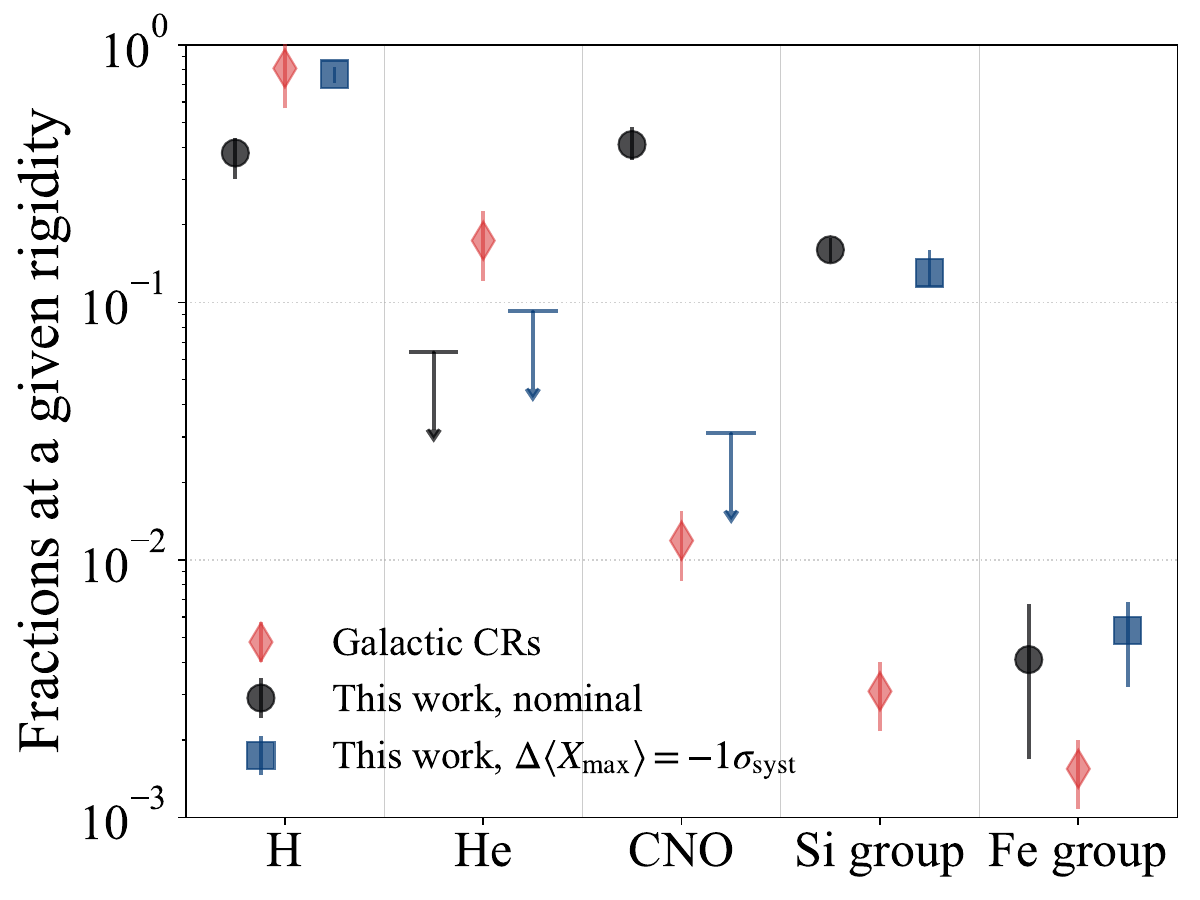}
    \caption{
    \justifying
        Comparison of 68\% credible intervals of fractions at a given \textit{rigidity} for UHECRs escaping from sources with the cosmological evolution following the star formation rate against values for Galactic cosmic rays. Black circles are for the fit with the nominal \textit{Auger} data, and the dark-blue squares are for the fit with the systematic shift of the \textit{Auger} data $\Delta \langle X_{\max} \rangle = -1\sigma_{\mathrm{syst}}$, resulting in the most conservative value of $Z_{\rho}$.
    }
    \label{fig:fractions_at_a_fixed_rigidity}
\end{figure}
The metallicity of GCRs to which we compare the UHECR metallicity is taken from Table 1 of Ref.\ \cite{Hoerandel:2002yg}. We assume a fixed universal index of GCRs of $\gamma_{\mathrm{GCR}} = 2.6$ when converting them to the fractions at a given \textit{rigidity} with Eq.~(\ref{eq:fractions_energy_to_rigidity}) (see the Supplemental Material for details). The relative uncertainty of these fractions is estimated to be around $\pm 30\%$ with similar values obtained for the difference between phenomenological models of the GCR spectra \texttt{polygonato}~\cite{Hoerandel:2002yg}, \texttt{GST}~\cite{Gaisser:2013bla}, \texttt{GSF-2025}~\cite{Dembinski:2025nmp,Fedynitch:2026ugq} and the GCR fluxes measured by the DAMPE collaboration~\cite{DAMPE:2025opn}. If UHECRs are GCR-like and reaccelerated by a rigidity-dependent process (as in e.g.~\cite{Caprioli:2015zka,Kimura:2017ubz}),
their fractions $\rho_i$ should match those of GCRs. This is the hypothesis we test.

\textit{Results}---We define the metallicity\footnote{This corresponds to the total \textit{number} of metal nuclei, i.e.\ elements heavier than helium, over the total number of nuclei in UHECRs (above a given \textit{rigidity}). Not to be confused with the metallicity defined sometimes as the \textit{mass} fraction of metals~\cite{MaiolinoandMannucci2019}.} of UHECRs as
\begin{equation}
    Z_{\rho} \equiv \rho_{\mathrm{N}} +\rho_{\mathrm{Si}} + \rho_{\mathrm{Fe}}.
\end{equation}
The $68\%$ credible intervals for $Z_{\rho}$ as a function of $m$ are shown in Fig.~\ref{fig:metallicity_as_function_of_m}. For all considered cosmological evolutions of UHECR sources, the inferred metallicity is at least a factor of ${\sim}3$ higher than the metallicity of GCRs.

Fits with $m \leq 0$ describe the data poorly (see the Supplemental Material), because the proton spectrum at Earth is then dominated by primary protons and is too narrow to match the observed one (see Discussion), whereas fits with $m > 0$ are better. For sources with cosmological evolution following the star formation rate, the discrepancy reaches a factor of ${\sim}4$ and becomes larger than an order of magnitude for scenarios with strongly positive source evolution with $m > 4$. This implies that either the CRs seeding UHECRs are more \textit{metal-rich} than GCRs, or that the reacceleration does not conserve their fractions $\rho_i$, i.e.\ it is not purely rigidity-dependent.

\begin{figure}[t]
    \centering
    \includegraphics[width=\linewidth]{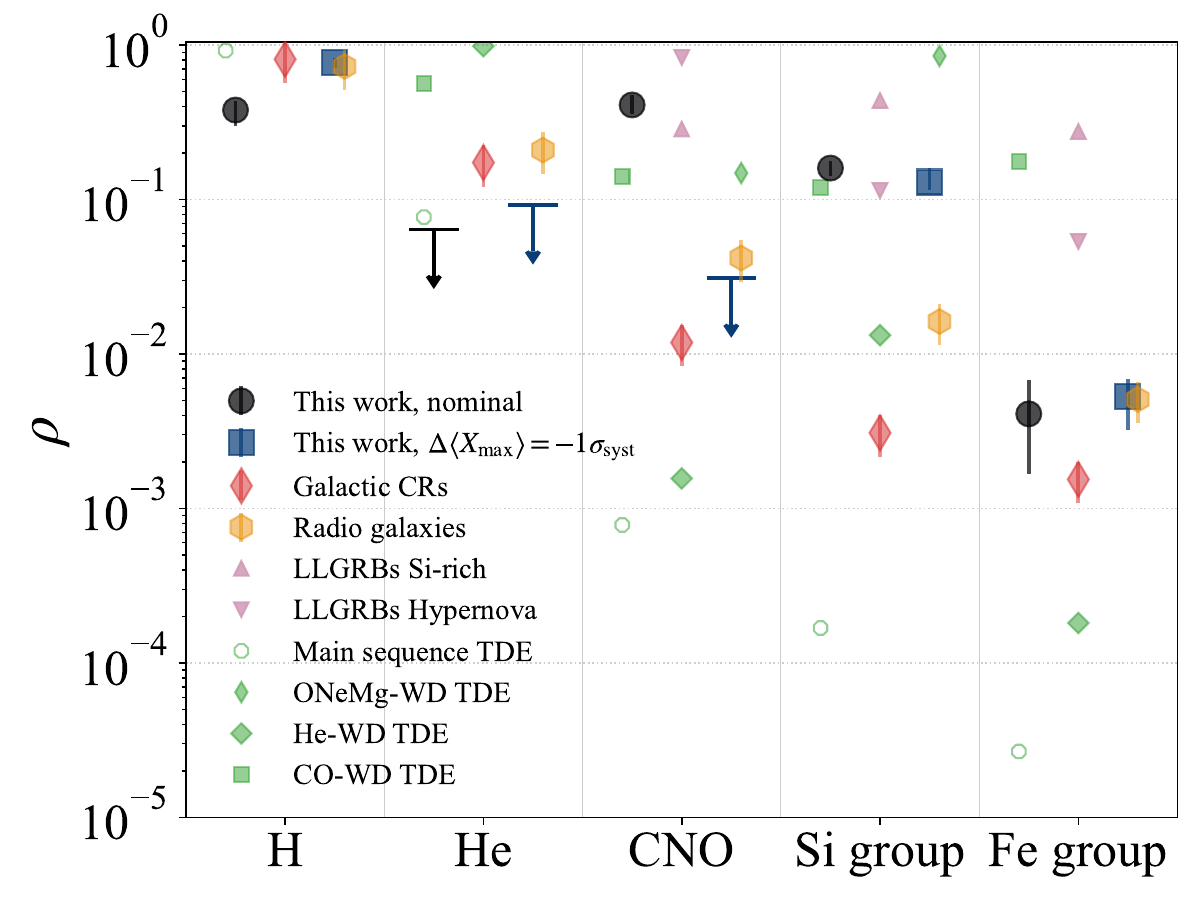}
    \caption{
    \justifying
        Same as Fig.~\ref{fig:fractions_at_a_fixed_rigidity} but with various model predictions from the literature.
    }
    \label{fig:theoretical_models}
\end{figure}

In Fig.~\ref{fig:metallicity_as_function_of_m}, we also plot estimated metallicities of UHECRs predicted by models of discrete shear acceleration of CRs in jets of radio galaxies~\cite{Kimura:2017ubz}, UHECR production in low-luminosity gamma-ray bursts (LLGRBs)~\cite{Zhang:2017moz, Woosley:2005gy} (and transrelativistic supernova \cite{Zhang:2018agl}), and tidal disruption events (TDEs) of white dwarfs (WD)~\cite{Zhang:2017hom, Rosswog:2008ie} (see also~\cite{Plotko:2024gop}). Models predicting too low a metallicity, namely reacceleration in radio galaxies and TDEs of He white dwarfs, cannot be reconciled with the \emph{Auger} data. Models injecting predominantly heavy nuclei, namely LLGRBs and TDEs of CO white dwarfs, overshoot $Z_{\rho}$ and would require substantial disintegration within the source (see also Fig.~\ref{fig:theoretical_models} for a broader comparison to phenomenological models in the literature).

A breakdown of the inferred fractions $\rho_{i}$ of UHECR nuclei at a given \textit{rigidity} is shown in Fig.~\ref{fig:fractions_at_a_fixed_rigidity}, where a source evolution following the star formation rate is assumed. We show credible intervals of $\rho_{i}$ for two cases: from our fit to the nominal \textit{Auger} data, and the variant with the shift $\Delta\langle X_{\max}\rangle = -1\sigma_{\mathrm{syst}}$, which allows, as explained below, the largest fraction of primary protons, hence the lowest metallicity. These cases differ significantly, with dominance either of CNO and Si groups, or Si and Fe groups in UHECRs above the ankle (see also the End Matter), which correspond to two qualitatively different descriptions of observed protons below the ankle. For the nominal \textit{Auger} data, the observed H flux below the ankle is described by secondary protons from the disintegration of CNO and Si nuclei. In this secondary-H-dominated solution, CNO and Si exceed their Galactic fractions by factors of ${\sim}30$ and ${\sim}50$, while escaping H is about half as abundant as in GCRs. For the primary-H-enriched solution with $\Delta \langle X_{\max} \rangle = -1\sigma_{\mathrm{syst}}$, primary and secondary protons contribute comparably below the ankle (with secondary protons coming mostly from disintegration of Si nuclei, while contributions of CNO and Fe groups are negligible). In the primary-H-enriched solution, H and CNO fractions are compatible with those in GCRs, but Si and Fe exceed their Galactic fractions by factors of ${\sim}50$ and ${\sim}3$. Counter-intuitively, although the shift $\Delta\langle X_{\max}\rangle = -1\sigma_{\mathrm{syst}}$ makes the composition observed at Earth heavier, it yields a \textit{lower} metallicity $Z_{\rho}$ of UHECRs escaping from the sources, due to a lower contribution of secondary nucleons from CNO.

The lack of primary He nuclei is explained by the high abundance of secondary He nuclei in UHECRs arriving at Earth due to photodisintegration of heavier primary nuclei during their intergalactic propagation. However, within hadronic interaction model uncertainties, primary helium at a level consistent with that of GCRs can be accommodated. The fitted all-particle and H spectrum, $\langle X_{\max} \rangle$, and $\sigma(X_{\max})$ for the nominal and shifted \textit{Auger} data are presented in the End Matter. The inferred high abundance of nuclei heavier than He at $E \gtrsim 5 \times 10^{18}$~eV is qualitatively consistent with previous works (see, e.g, Refs.~\cite{Ehlert:2023btz, Ehlert:2022jmy, PierreAuger:2016use, PierreAuger:2023htc, Plotko:2022urd, Luce:2022awd, Zhang:2024sjp, Alhebsi:2026bdk, Marafico:2024qgh, Heinze:2020zqb}, but see also~\cite{Strasman:2026ech}).

\textit{Discussion}\\
\indent\textit{Composition systematics}---The dominant uncertainty of this study is due to the inference of the composition from air-shower data. These originate from
uncertainties in the $X_{\max}$ scale arising from both experimental uncertainties and the predictions of hadronic interaction models. Our fiducial analysis uses \EPOSLHCR and the nominal \textit{Auger} data. The experimental uncertainty is $\pm 10\,\mathrm{g/cm^2}$, while \SibyllTwoThreeE predicts $X_{\max}$ values differing by about $-10\,\mathrm{g/cm^2}$.  We repeat the analysis with both models, shifting the data by $\pm 1$ experimental systematic uncertainty. Relative to the fiducial case, this effectively covers shifts from $-20$ to $+10\,\mathrm{g/cm^2}$. We also consider the more extreme shift in the $X_{\max}$ scale of Ref.~\cite{Vicha:2025fxx}, corresponding to approximately $+20\,\mathrm{g/cm^2}$ relative to our fiducial case. Hence the range of $X_{\max}$ scale explored in this work corresponds to about $40\%$ of the overall difference of proton- and iron-induced air showers, and can therefore be considered conservative.

\begin{figure*}[h]
 \panel{4plots_lin_log_EPOS-LHC-R---CENTRE_GROUND_noLabel}{a}\panel{4plots_lin_log_EPOS-LHC-R---CENTRE_DOWN_noLabel}{b}
\panel[12,59]{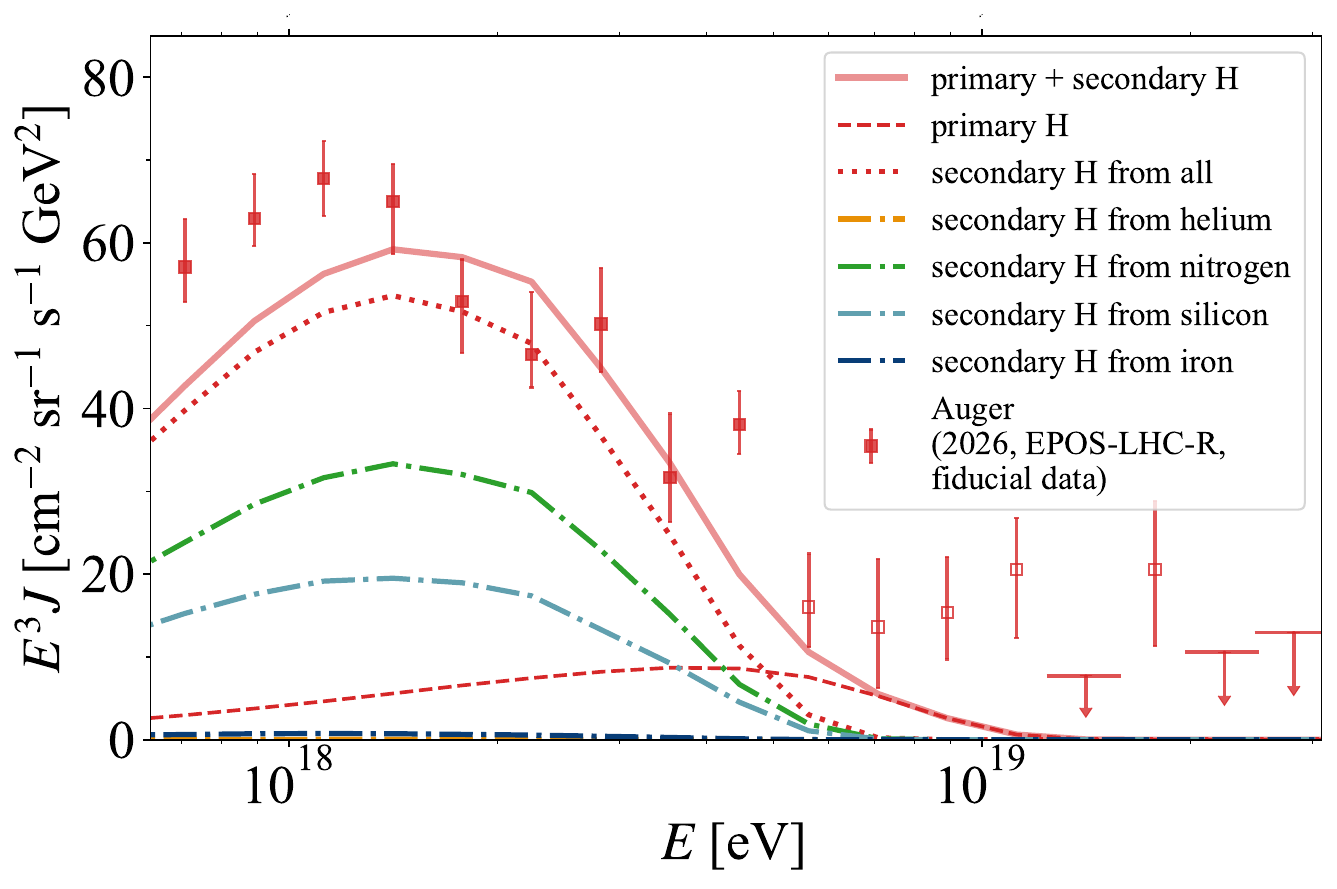}{c}\panel[12,59]{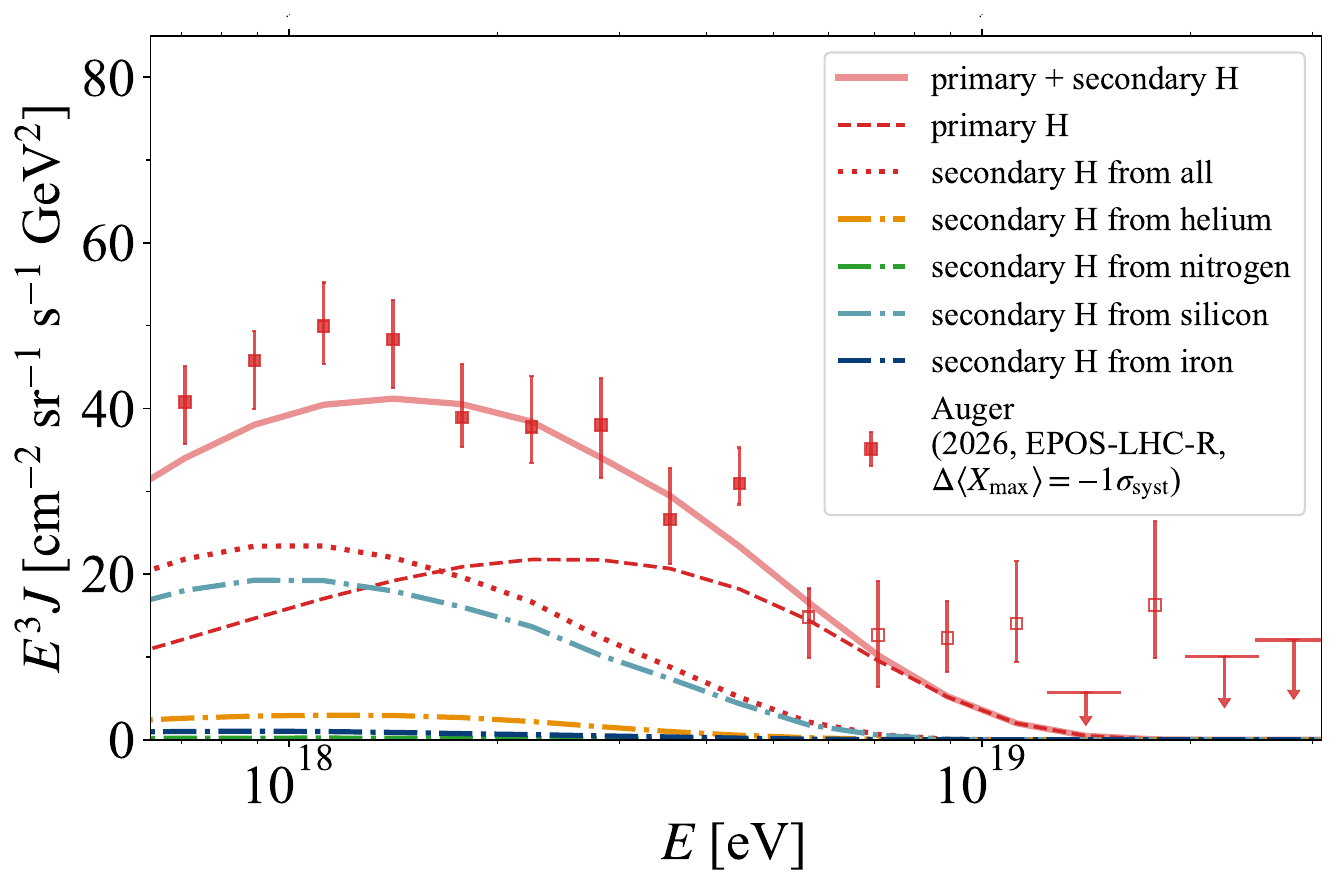}{d}
\panel[15,66]{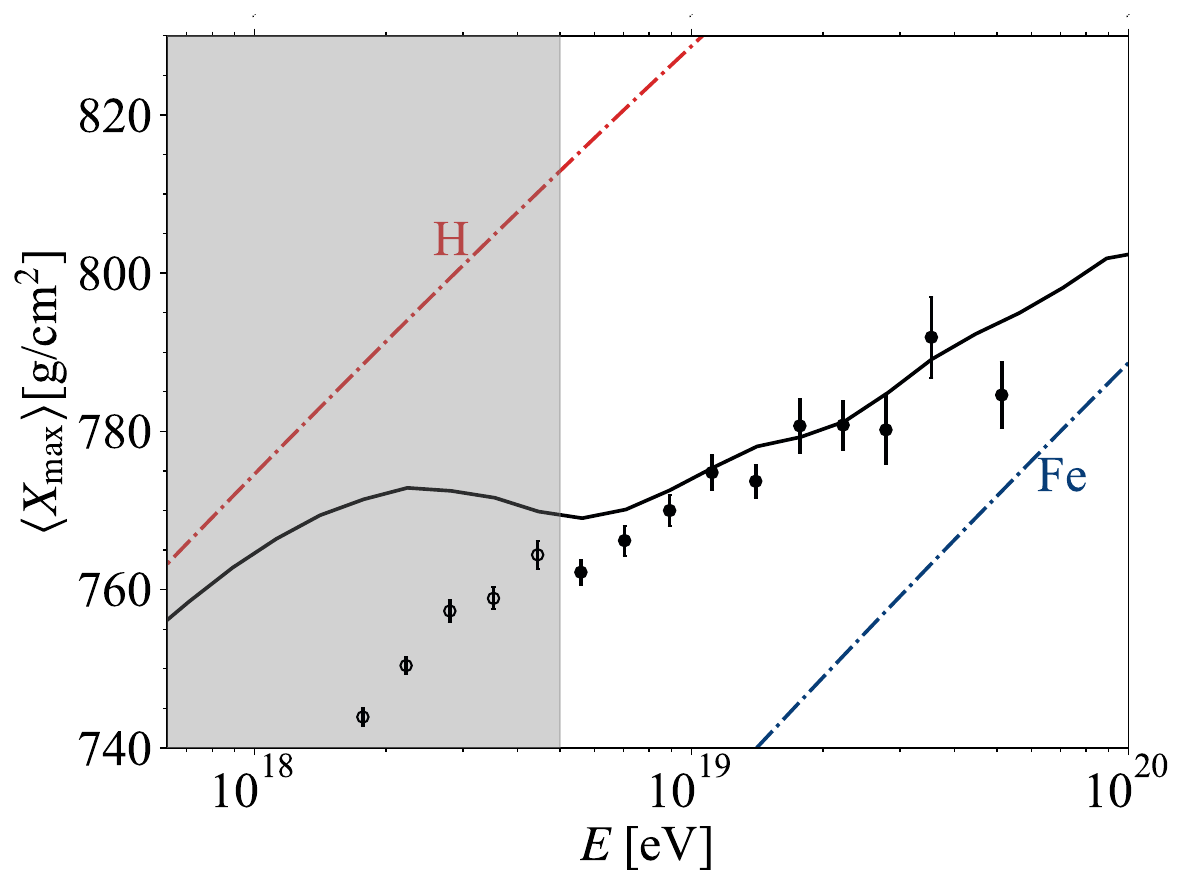}{e}\panel[15,66]{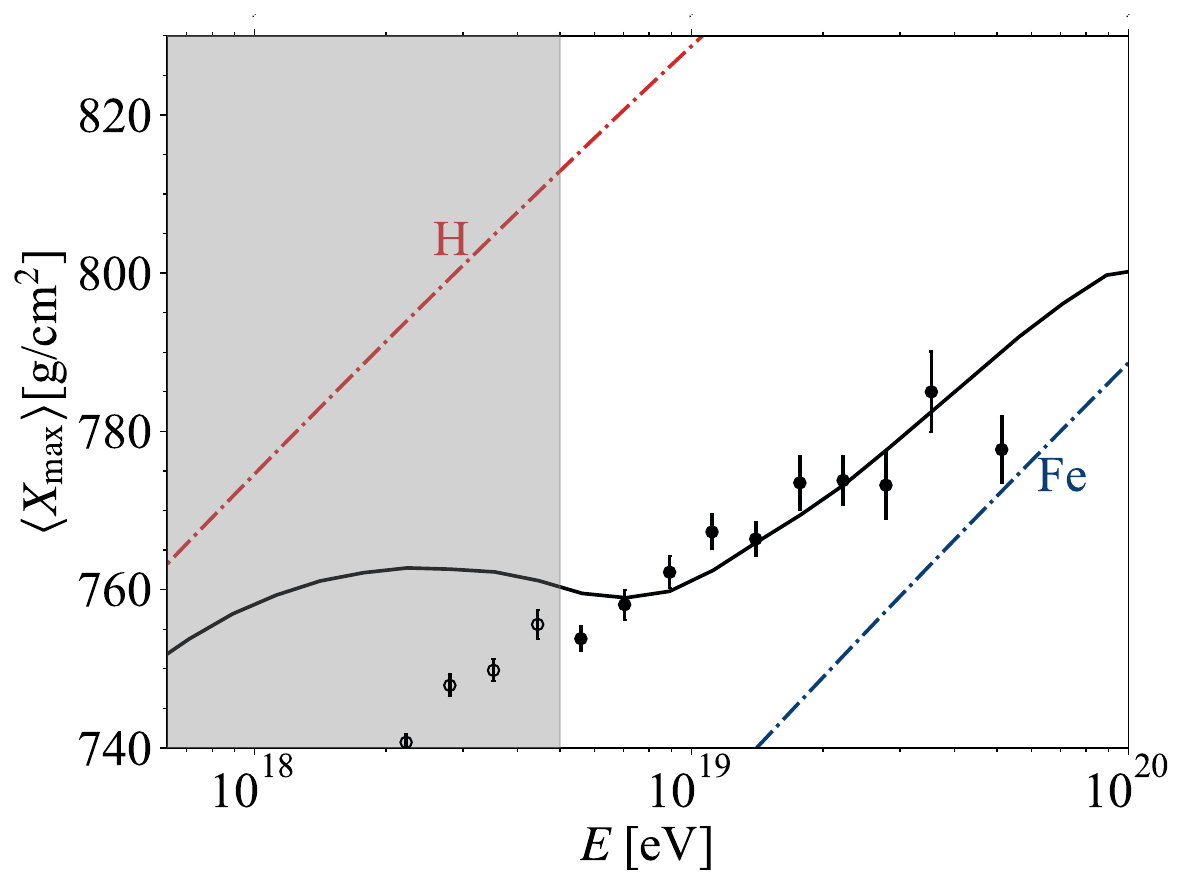}{f}
\panel[13,66]{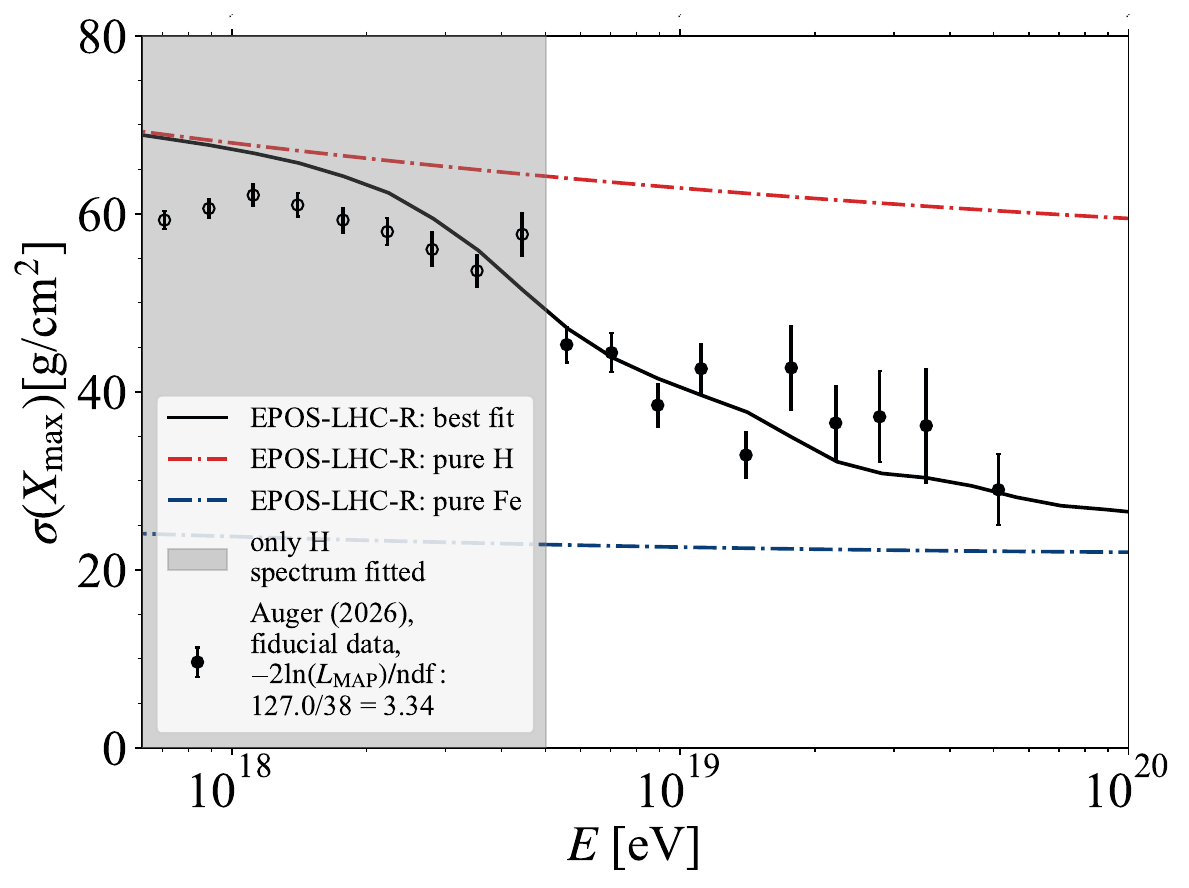}{g}\panel[13,66]{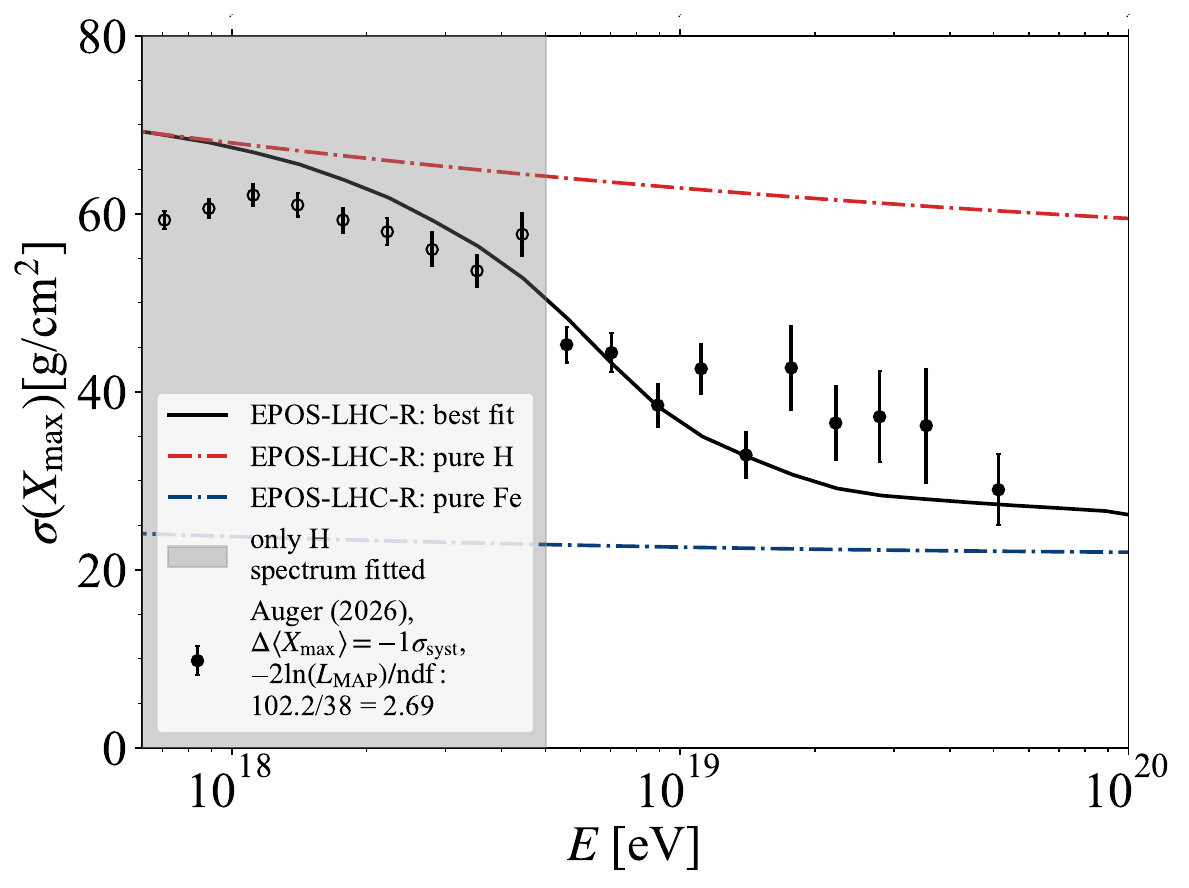}{h}
  
    \caption{
    \justifying
        All-particle and H spectra and first two moments of the $X_{\max}$ distribution calculated with our model (curves) and data reported by \textit{Auger} (circles with error bars for the spectrum and the $X_{\max}$ moments, squares for the H spectrum; hollow markers are not included in the likelihood function).
        Panels on the left: the secondary-H-dominated scenario; panels on the right: the primary-H-enriched scenario. Panels (a)-(b): intensity times cosmic-ray energy cubed. Panels (c)-(d): the proton spectrum at Earth with a breakdown of the secondary proton flux from disintegration of different primary heavier nuclei shown in dash-dotted lines according to the legend. Panels (e)-(f): the mean depth of the air-shower maximum vs. energy; dash-dotted lines show predictions of EPOS-LHC-R for beams of pure hydrogen and iron nuclei. Panels (g)-(h): the standard deviation of the depth of the air-shower maximum vs. energy. The upper limits for the all-particle spectrum are at the 90\% confidence level, while for the H spectrum at the 68\% confidence level.
    }
    \label{fig:8plots}
\end{figure*}

In the secondary-H-dominated scenario, the inferred composition at Earth is relatively light, but the source metallicity is high because most protons
originate from CNO photodisintegration (cf.~Fig.~\ref{fig:8plots}(c) in the End Matter) with some contribution from disintegration of Si nuclei as well. Shifting the $X_{\max}$ scale toward heavier compositions suppresses the CNO contribution, reduces the source metallicity and leads to the primary-H-enriched scenario, as primary silicon becomes the dominant source of secondary protons (cf.~Fig.~\ref{fig:8plots}(d) in the End Matter). For even heavier compositions, the proton fraction at Earth decreases further, again increasing the inferred source metallicity. Thus, despite these variations, a strong enhancement of heavy nuclei at the sources is required in all cases.

At present, the two scenarios cannot be distinguished within the
experimental and theoretical uncertainties of the $X_\mathrm{max}$
scale. Future composition measurements after the \textit{AugerPrime} upgrade~\cite{Castellina:2019irv}
together with improved hadronic-interaction models constrained by proton–oxygen
collisions at the LHC (e.g.\,\cite{ATLAS:2026xcm,Adriani:2026ses}) and
pion–air interactions constrained by muon production depth
measurements~\cite{Ostapchenko:2016bir}, may resolve this ambiguity.

\textit{Robustness of the comparison}---The metal excess we infer is robust to the main assumptions entering the comparison. For seed populations such as GCRs, conservation of fractions at fixed rigidity ($\rho_i$) requires
a common spectral index, a rigidity-dependent cutoff,
and a reacceleration that conserves the number of each species above a rigidity threshold.
In-source photodisintegration breaks these conditions. Neutrons, being unconfined by magnetic fields,
escape with a softer spectrum~\cite{Baerwald:2013pu},
while nuclei escape with hardened spectra~\cite{Globus:2015xga, Unger:2015laa}.
However, photodisintegration will always \textit{decrease} the metallicity escaping from the source relative to the injected one.
Therefore, the material injected in the sources must be at least as metal-rich as we infer,
which strengthens the tension with GCRs.

\textit{Source evolution}---The fitted spectral index, and hence the inferred $\rho_i$, depends on the propagation scenario:
a dominant contribution from nearby sources~\cite{Bister:2023icg, Bister:2024ocm, Bister:2025emc, Mollerach:2024sjd}
or a magnetic horizon~\cite{Mollerach:2024sjd, PierreAuger:2024hlp, Cermenati:2026qny}
would soften the inferred source spectra~\cite{Caprioli:2015zka, Taylor:2015rla}
and reduce the metal excess. However, even for sources with strongly negative redshift evolution,
which yield the softest fitted source spectra,
the inferred metallicity $Z_{\rho} \gtrsim 0.08$ exceeds that of GCRs
($Z_{\rho} \lesssim 0.02$) by at least a factor of ${\sim}3$ (cf.\ Fig.~\ref{fig:metallicity_as_function_of_m}).

Fits with negative source evolution are, however, poorer. After intergalactic propagation, the Peters cycle consists of $N+1$ cycles:
$N$ mass groups with maximum energies $Z_i R_{\max}$, and a cycle of secondary protons, which carry $E/A$ of their parent nuclei.
The maximum energy of the latter is about half that of the primary protons,
\begin{equation}
    E_{\max}^{\mathrm{sec}} = R_{\max}\, Z/A \approx R_{\max}/2.
\end{equation}

The observed proton spectrum is best described by a mixture of the two components, whereas scenarios yielding mostly primary protons (negative source evolution)
or mostly secondary protons (the fiducial CNO-dominated solution) produce a proton spectrum that is too narrow. UHECRs from nearby sources alone, which would yield few secondary protons,
are therefore disfavoured by the observed proton spectrum. More stringent observational tests on the broadness of the proton distribution, and thus on the number of proton components (secondary or primary or both), will be possible with low-energy measurements of $X_{\max}$ from HEAT~\cite{Mathes:2011zz} and TALE~\cite{TelescopeArray:2026rdu}.

\textit{Candidate sources}---The hydrogen deficit of the fiducial scenario points to sources depleted of hydrogen,
such as stripped stellar progenitors or white dwarfs, but it is not robust to the composition systematics.
On the contrary, the Si--Fe excess is. While a Si component improves our fits, other analyses of the same data do not require it~\cite{PierreAuger:2026qbt},
and we therefore consider Si and Fe together in what follows.
Even when the H abundance matches that of GCRs, UHECR sources must be strongly enriched in the Si--Fe group.

Radio galaxies~\cite{Kimura:2017ubz} and AGN outflows~\cite{Ehlert:2024lji},
which reaccelerate ambient CRs, could reproduce the inferred composition
only if their seed CRs were enriched in silicon/iron by a factor of ${\sim}30$ relative to GCRs (cf. Fig.~\ref{fig:theoretical_models} in the End Matter)\footnote{If the lower-energy seeds in the hosts have harder spectra than GCRs, then the heavy-element excess can be smaller.}. Preferential injection of weakly ionised, high mass-to-charge ions~\cite{Caprioli:2017oun, Hanusch:2018bsk, Caprioli:2025lor} and of refractory elements locked in dust grains, which already shapes the GCR composition~\cite{Epstein1980MNRAS, Meyer:1997vz, Ellison:1997an, Eichmann:2020dtn}, would have to be far more efficient than in the sources of GCRs.

TDEs can supply enough metals only for CO or ONe white dwarf disruptions,
which are too rare to power the bulk of UHECRs~\cite{Zhang:2017hom, Plotko:2024gop}
(but see~\cite{Biehl:2017hnb}). High-luminosity GRBs, originally proposed as sources of UHE protons~\cite{Waxman:1995vg,Vietri:1995hs}, may also synthesise heavier nuclei~\cite{Metzger:2011xs}.
However, they disintegrate nuclei at high radiation densities~\cite{Biehl:2017zlw,Zhang:2017moz}.
Multi-collision models can describe the data, but only for a fitted, metal-rich injection composition ($Z_{\rho} \approx 0.7$~\cite{Heinze:2020zqb}), whose origin in GRB progenitors or outflows remains to be confirmed. LLGRBs are less well understood as a population. Jets loaded with material from the stripped progenitor~\cite{Zhang:2017moz}, for any of the progenitor compositions considered there,
could reproduce the inferred composition, given the right amount of in-source photodisintegration (cf.~\cite{Globus:2014fka, Globus:2015xga, Boncioli:2018lrv}) or admixture of lighter material from the central engine or circumstellar environment. In this simplified comparison, we assume that the accelerated number fractions of heavy elements are the same as those of the progenitor's material, which need not be the case. For binary neutron star mergers~\cite{Farrar:2024zsm,Farrar:2025kpk,Suliga:2026vys}, rich in $r$-process material, no \textit{a priori} abundances are available.

\textit{Conclusion}---We have shown that the elemental composition of UHECRs
must be compared with candidate seed populations \emph{at a given rigidity}.
Abundances quoted at a given energy understate the excess of heavy nuclei
when populations with different spectral indices, such as soft GCRs and hard UHECRs, are compared.
We also showed that, at a given rigidity, UHECRs escaping their sources are strongly enriched in heavy nuclei,
by more than an order of magnitude relative to Galactic cosmic rays in the silicon-iron group.
This conclusion holds for all source evolutions, $X_{\max}$ systematics,
and hadronic interaction models considered.

Simple reacceleration of GCR-like seeds, as in radio-galaxy jets, is therefore disfavoured.
None of the source models with a-priori composition reproduces the inferred composition as published.
The composition instead seems consistent with metal-rich debris,
accelerated where radiation fields are weak enough for nuclei to survive.
Thus, the origin of the heavy-element excess of UHECRs remains an open question.
% ----------------------------------------------------------------------
% ACKNOWLEDGMENTS
% ----------------------------------------------------------------------
\begin{acknowledgments}
\textit{Acknowledgments}---We thank Damiano Caprioli, Kohta Murase, Anatoli Spitkovsky, Tsvi Piran, Bing Theodore Zhang, Walter Winter, Igor Vaiman, Maksat Satybaldiev, Bidisha Sen, and Mathias Pavely Nødtvedt for useful discussions.

This work made use of \texttt{emcee}~\cite{emcee-Foreman-Mackey-2013},
\texttt{ArviZ}~\cite{arviz_2019}, \texttt{crflux}~\cite{Fedynitch:2012fs},
\texttt{numpy}~\cite{numpy}, \texttt{scipy}~\cite{2020SciPy-NMeth},
\texttt{matplotlib}~\cite{Hunter:2007},
\texttt{snakemake} \cite{Moelder2021},
\texttt{corner.py} \cite{corner-Foreman-Mackey-2016},
and
\texttt{astropy}~\cite{astropy:2013,astropy:2018,astropy:2022}.

\end{acknowledgments}

\textit{Author Contributions}---E.P.\ wrote scripts for fitting the models to the data, produced all the figures and tables and prepared the original draft; F.O.\ and M.U.\ conceived the study and supervised the project at all stages; D.E.\ performed the \texttt{CRPropa3} simulations. All the authors reviewed and edited the manuscript.
% ----------------------------------------------------------------------
% DATA AND CODE AVAILABILITY
% ----------------------------------------------------------------------

\textit{Data and code availability}---The data used in this work are publicly available at \cite{PierreAuger:2025eun, PierreAuger:2026qbt, AugerZENODO2026}. The code reproducing the results presented in the paper will be published on \textit{Zenodo} upon acceptance of the manuscript.

\textit{Use of AI}---Large language models were used for minor language editing and to assist with writing selected code snippets, which were subsequently tested and verified by the authors.

% ----------------------------------------------------------------------
% BIBLIOGRAPHY
% ----------------------------------------------------------------------
\bibliographystyle{apsrev4-2}   % default RevTeX style; change if you prefer another
\bibliography{biblio}          % <-- put your *.bib file name here (e.g. biblio.bib)

% ----------------------------------------------------------------------
% END MATTER
% ----------------------------------------------------------------------
\cleardoublepage
\begin{center}
\textbf{End Matter}
\end{center}
\twocolumngrid
Here, we present a detailed comparison of our fractions at a given rigidity with phenomenological models present in the literature (Fig.~\ref{fig:theoretical_models}), as well as the results for the combined fit to the \textit{Auger} data for the two solutions discussed in the main text (Fig.~\ref{fig:8plots}).
In Fig.~\ref{fig:theoretical_models}, we compare our fitted fractions
(as in Fig.~\ref{fig:fractions_at_a_fixed_rigidity}) with the fractions $\rho_i$ at fixed rigidity of H, He, CNO, Si, and Fe groups predicted by models of discrete shear acceleration of CRs in jets of radio galaxies~\cite{Kimura:2017ubz}, UHECR production in low-luminosity gamma-ray bursts (LLGRBs)~\cite{Zhang:2017moz, Woosley:2005gy}, and tidal disruption events (TDEs) of white dwarfs (WD)~\cite{Zhang:2017hom, Rosswog:2008ie}. The radio-galaxy model~\cite{Kimura:2017ubz} reaccelerates seed CRs assuming metals enriched by a factor of ${\sim}3$ relative to GCRs. It overshoots our upper limits on He and falls short of the fitted Si fractions in both solutions. The LLGRB models~\cite{Zhang:2017moz} are qualitatively similar for all progenitor compositions considered. Transrelativistic supernovae are also bracketed by these progenitor models, as they are expected to have the same injected composition (see~\cite{Zhang:2018agl}). We show models for a Si-rich LLGRB and a hypernova, but have checked all six of the models presented in~\cite{Zhang:2017moz}. They contain no H or He and therefore fall short of the fitted H fractions, while they overshoot the fitted Fe fractions in both solutions. Note that since we work only with five element groups, we assign S nuclei to the Fe group. As mentioned above, additional in-source interactions could reconcile these models with the observed UHECR composition fractions. Among the TDE models~\cite{Zhang:2017hom}, disruptions of He white dwarfs overshoot our upper limits on He and fall short of the fitted metal fractions, along with TDEs of main-sequence stars, which have ${\sim}10^{2}$ less metals than in our fit. Disruptions of ONeMg and CO white dwarfs contain too few protons. The ONeMg model overshoots the silicon group, and the CO model contains more metals than required. In-source photodisintegration could in principle bring the latter into agreement with our fits, but such events are likely too rare to power the bulk of the UHECR flux.

Fig.~\ref{fig:8plots} shows the fits for the secondary-H-dominated and primary-H-enriched scenarios respectively. The former solution (panels on the left in Fig.~\ref{fig:8plots}) results in the dominance of the secondary protons over the primary ones in the whole fitted energy range, with the major contribution of secondary nucleons from disintegration of CNO nuclei during their intergalactic propagation and a smaller contribution from the Si group. The latter solution (panels on the right in Fig.~\ref{fig:8plots}) has a significantly lower CNO flux, which leaves only the Si group of primary nuclei to produce secondary protons. The lowered observed flux of protons due to the shift $\Delta \langle X_{\max} \rangle = -1\sigma_{\mathrm{syst}}$ allows an increase in flux of primary protons without a significant worsening of the fit to the all-particle spectrum $J(E)$ and $\langle X_{\max} \rangle$. Thus, the second solution results in a lower (more conservative) value of metallicity $Z_{\rho}$ of UHECRs.
\clearpage
\begin{center}
\textbf{\large Supplemental Material for
``Ultra-high-energy cosmic rays require metal-rich sources''}
\end{center}
\vspace{1em}
\twocolumngrid
\setcounter{section}{0}
\makeatletter
\patchcmd{\@ssect@ltx}
    {\addcontentsline{toc}{#1}{\protect\numberline{}#8}}
    {}
    {}
    {}
\makeatother
\definecolor{rossoCP3}{cmyk}{0,.88,.77,.40}
\definecolor{darkBlue}{rgb}{0, 0, 0.8}
\hypersetup{
  bookmarks=true,         % show bookmarks bar?
  bookmarksopen=true,     % open bookmarks
  bookmarksopenlevel=1,
  unicode=false,          % non-Latin characters in Acrobat’s bookmarks
  pdftoolbar=true,        % show Acrobat’s toolbar?
  pdfmenubar=true,        % show Acrobat’s menu?
  colorlinks=true,        % false: boxed links; true: colored links
  linkcolor=darkBlue,     % color of internal links
  citecolor=darkBlue,     % color of links to bibliography
  filecolor=darkBlue,     % color of file links
  urlcolor=darkBlue       % color of external links
}

% This block is written by Claude Sonnet 5
\renewcommand{\theequation}{S\arabic{equation}}
\renewcommand{\thefigure}{S\arabic{figure}}
\renewcommand{\thetable}{S\arabic{table}}
\setcounter{equation}{0}
\setcounter{figure}{0}
\setcounter{table}{0}
% This block is written by Claude Sonnet 5

% This block is written by Rool
\makeatletter
\def\@hangfrom@section#1#2#3{\@hangfrom{#1#2}#3}%  % numbered sections
\def\@hangfroms@section#1#2{#1#2}%                 % starred sections
\makeatother
% This block is written by Rool

\section{Cosmic-Ray Elemental Abundances After Reacceleration}\label{sec:terminology}
Let us assume that CRs have a spectrum described by the following equation:
\begin{equation}
    \left(\frac{dN}{dE} \right)_{i,\alpha} = C_{i,\alpha} \left( \frac{E}{E_{\mathrm{ref}}} \right)^{-\gamma_{\alpha}} \exp \left[- \left( \frac{E}{E^{\mathrm{cut}}_{i, \alpha}} \right)^{\lambda} \right].
    \label{eq:S_spectrum}
\end{equation}
In the beginning ($\alpha = 1$), CRs have spectral index $\gamma_{1}$ and cutoff energy $E^{\mathrm{cut}}_{i, 1}$. Then they undergo some process of reacceleration ($\alpha = 2$), which makes their spectral index equal to $\gamma_{2}$ and boosts their cutoff energy to $E^{\mathrm{cut}}_{i, 2}$.

How the relative abundances of CRs of type $i$ change after reacceleration depends on the details of this process, but, in the simplest case, we can require that this process does not change the total number $N^{\mathrm{tot\,}}_{i, \alpha}$ of CRs of any type $i$, i.e.\ CRs do not disintegrate and are not produced by external particles. The total number of CRs can be expressed as
\begin{multline}
    N^{\mathrm{tot}}_{i, \alpha} = \int \limits_{E^{\min}_{i, \alpha}}^{+\infty} \left(\frac{\dd N}{\dd E} \right)_{i, \alpha} \dd E = \\ = \frac{C_{i, \alpha} E_{\mathrm{ref}}^{\gamma_{\alpha}} \left( E^{\mathrm{cut}}_{i, \alpha} \right)^{1-\gamma_{\alpha}}}{\lambda} \Gamma\left[ \frac{1 - \gamma_{\alpha}}{\lambda}, \left ( \frac{E^{\min}_{i, \alpha}}{E^{\mathrm{cut}}_{i, \alpha}} \right)^{\lambda} \right],
\label{eq:S_total_number_of_nuclei}
\end{multline}
where $\Gamma$ is the upper incomplete gamma function. Conservation of the total number of nuclei $i$ means $N^{\mathrm{tot}}_{i,1} = N^{\mathrm{tot}}_{i, 2}$. Thus, we obtain, using Eq.~(\ref{eq:S_total_number_of_nuclei}), the normalisation $C_{i, 2}$ of nuclei $i$ at a given \textit{energy} after reacceleration:
\begin{equation}
    C_{i, 2} = C_{i, 1} \cdot \frac{(E^{\mathrm{cut}}_{i, 1})^{1 - \gamma_{1}}}{(E^{\mathrm{cut}}_{i, 2})^{1 - \gamma_{2} }} \cdot \frac{\Gamma\left[ \frac{1 - \gamma_{1}}{\lambda}, \left ( \frac{E^{\min}_{i, 1}}{E^{\mathrm{cut}}_{i, 1}} \right)^{\lambda} \right]}{\Gamma\left[ \frac{1 - \gamma_{2}}{\lambda}, \left ( \frac{E^{\min}_{i, 2}}{E^{\mathrm{cut}}_{i, 2}} \right)^{\lambda} \right]}.
\label{eq:S_normalisation_after_reacceleration}
\end{equation}

\subsubsection{Conservation of the total number of nuclei above a given \textit{energy}}

Consider first the case in which the energy of the CRs after reacceleration increases by a constant factor $\beta$, and the minimum and cutoff energies are the same for any type $i$ of nuclei, i.e. $E^{\min}_{i, \alpha} = E^{\min}_{\alpha}$, $E^{\mathrm{cut}}_{i, \alpha} = E^{\mathrm{cut}}_{\alpha}$ ($\alpha = 1,2$). Then for two different species of nuclei $i$ and $j$ with charges $Z_{i} > Z_{j}$, their fractions at a given \textit{energy} and the ratio of the abundances of elements $i$ and $j$, $C_{i, \alpha} / C_{j, \alpha} = \varepsilon_{i, \alpha} / \varepsilon_{j, \alpha}$, after reacceleration will not change, because in this case there is no dependence on $Z$ in Eq.~(\ref{eq:S_normalisation_after_reacceleration}).

\subsubsection{Conservation of the total number of nuclei above a given \textit{rigidity}}

If, instead, the minimum \textit{rigidity} is universal for all types $i$ of nuclei, and there is a \textit{rigidity}-dependent cutoff in their spectrum, then $E^{\min}_{i, \alpha} = Z_{i} R^{\min}_{\alpha}$, $E^{\mathrm{cut}}_{i, \alpha} = Z_{i} R^{\mathrm{cut}}_{\alpha}$ ($\alpha = 1,2$). In this case, abundances at a given \textit{rigidity} will not change, and, on the contrary, abundances at a given \textit{energy} will change if $\gamma_2 \neq \gamma_1$:
\begin{equation}
    \frac{\varepsilon_{i, 2}}{\varepsilon_{j, 2}} = \frac{\varepsilon_{i, 1}}{\varepsilon_{j, 1}} \left( \frac{Z_{i}}{Z_{j}} \right)^{\gamma_2 - \gamma_1}.
    \label{eq:S_change_of_metallicity}
\end{equation}

Thus, the answer to the question ``How does the metallicity of UHECRs change after the reacceleration?'' is that for any $\gamma_{1}$ and $\gamma_{2}$, if CRs are reaccelerated by a \textit{rigidity}-dependent process, and the total number of each species of CRs is conserved, then at a given \textit{energy}, CRs will appear with \textit{higher metallicity} if the reaccelerated spectrum is \textit{softer} ($\gamma_{2} > \gamma_{1}$), and with \textit{lower metallicity} if the reaccelerated spectrum is \textit{harder} ($\gamma_{2} < \gamma_{1}$), while metallicity of CRs at a given \textit{rigidity} remains the same.

Source abundances are often quoted at a fixed energy (e.g.,~\cite{PierreAuger:2016use}) or as a fraction of the total energy carried by a given nucleus type above a fixed energy threshold (e.g.,~\cite{Plotko:2022urd}). The former convention is very sensitive to differences in spectral indices before and after reacceleration (see Fig.\,\ref{fig:schematic}), thus, can underestimate the heavy-metal excess of hard-spectrum UHECRs with respect to the soft-spectrum GCRs. The latter is not that sensitive to $\gamma$ as long as the spectra have a rigidity-dependent cutoff and are hard enough ($\gamma < 2$), so that the total CR energy is dominated by the high-energy rigidity-dependent tail. Otherwise, the inferred fractions will depend on the ratio of nucleus spectra at the threshold energy, which falls back to estimating fractions at a fixed energy. Our proposed convention with fractions $\rho_{i}$ defined in Eq.\,(\ref{eq:fractions_energy_to_rigidity}) is valid for any value of $\gamma$ as long as the spectra of all nuclei share the same spectral index, a rigidity-dependent cutoff, and a reacceleration that conserves the number of each species above a rigidity threshold.

\section{Intergalactic Propagation of UHECRs}\label{sec:propagation}
The sources of UHECRs are simulated in the continuous-source approximation from a minimum cosmological redshift $z_{\min} = 0.001$ out to a maximum redshift $z_{\max} = 4$. The UHECR sources are simulated to emit five types of nuclei: hydrogen H-1, helium He-4, nitrogen N-14, silicon Si-28, and iron Fe-56. UHECRs are propagated in the intergalactic space, taking into account adiabatic redshift energy losses, photopion production, Bethe-Heitler pair production, and photodisintegration of nuclei on the Cosmic Microwave Background (CMB) and the Extragalactic Background Light (EBL).

\section{Spectrum of UHECRs Escaping from Sources and the Fitting Procedure}\label{sec:fitting}
In our approach of a continuous source distribution, we assume that UHECRs escape from each source with an energy spectrum defined as a power law with a \textit{rigidity}-dependent (``Peters cycle'' \cite{Peters:1961mxb}) superexponential cutoff (see Eq.~\ref{eq:S_spectrum}), where $E_{\mathrm{ref}} = 10^{18}$~eV (fixed), $E^{\mathrm{cut}}_{i} = Z_{i} R_{\mathrm{cut}}$, the rigidity cutoff $R_{\mathrm{cut}}$ is a free parameter along with the spectral index $\gamma$ and the power of the superexponential cutoff $\lambda$. Four (for $i = $ He, N, Si, Fe) fractions at a given energy are free parameters, while the fifth fraction of hydrogen is calculated as $\varepsilon_{\mathrm{H}} = 1 - \varepsilon_{\mathrm{He}} - \varepsilon_{\mathrm{N}} - \varepsilon_{\mathrm{Si}} - \varepsilon_{\mathrm{Fe}}$. The total normalisation of the UHECR spectrum $C_{\mathrm{tot}} = \sum_i C_{i}$ is the last free parameter. Thus, our model has eight free parameters.
    
To fit our model parameters to the measured UHECR spectrum, $\langle X_{\max}\rangle$ and $\sigma\left( X_{\max} \right)$, we perform the MCMC inference using the \texttt{emcee} package \cite{emcee-Foreman-Mackey-2013}. We run 32 MCMC walkers over the seven free parameters $\gamma$, $R_{\mathrm{cut}}$, $\lambda$, $\varepsilon_{\mathrm{He}}$, $\varepsilon_{\mathrm{N}}$, $\varepsilon_{\mathrm{Si}}$, and $\varepsilon_{\mathrm{Fe}}$, while the normalisation parameter $C_{\mathrm{tot}}$ is not a part of the MCMC chain and is derived analytically via the method of least squares in each step of the chain.

We use \textit{uninformative} priors for all the free parameters with the following bounds: $-2.4 \leq \gamma \leq 2.6$, which corresponds to the spectral index hardened with respect to that of GCRs by $\Delta_{\gamma} \in [0; 5]$; $5 \times 10^{16}$~V $\leq R_{\mathrm{cut}} \leq$ $1 \times 10^{20}$~V, which corresponds to the boost of the rigidity cutoff of $5$~PV of GCRs by a factor of $\beta \in [10; 20\,000]$; $ 0.5 \leq \lambda \leq 15$. As for $\varepsilon_{\mathrm{He}}$, $\varepsilon_{\mathrm{N}}$, $\varepsilon_{\mathrm{Si}}$, $\varepsilon_{\mathrm{Fe}}$, when they are added together with the dependent $\varepsilon_{\mathrm{H}}$, they sum up to unity. Accepting the next step of the MCMC if each of the four free fractions is within $[0; 1]$, defining the dependent hydrogen fraction as
    \begin{equation}
        \varepsilon_{\mathrm{H}} = \max \left\{0; 1 - \varepsilon_{\mathrm{He}} - \varepsilon_{\mathrm{N}} - \varepsilon_{\mathrm{Si}} - \varepsilon_{\mathrm{Fe}} \right\},
    \end{equation}
and demanding that the sum of the five fractions is unity, effectively results in the \textit{uninformative} \textit{joint} prior Dirichlet distribution, uniform on the $(5-1)$ simplex, where the \textit{marginal} prior distribution of each of the five fractions is a Beta$(1;4)$ distribution. We also did a test fit with an \textit{informative} prior with the \textit{marginal} prior distribution of each of the five fractions distributed as Beta$(5;20)$, which provided similar posterior credible intervals of the model parameters.

The estimation of the likelihood function relies on the following data: (1) all-particle spectrum $J(E)$, where intensity $J(E) \equiv {\dd N}/{(\dd E \, \dd t \, \dd A \, \dd \Omega)}$, $\dd t$ is an element of time, $\dd A$ is an element of surface area, $\dd \Omega$ is an element of solid angle; (2) $\langle X_{\max} \rangle$ and (3) $\sigma\left( X_{\max} \right)$ as functions of energy; (4) the H spectrum, which is the product of the \textit{Auger} ``proton (energy-dependent) fractional abundances'' from Refs.~\cite{PierreAuger:2026qbt, AugerZENODO2026}\footnote{These abundances depend on the assumed HIM and the $X_{\max}$ scale. We use posterior distributions inferred from the published posterior samples of MCMCs in \cite{AugerZENODO2026} for each HIM and shift in $\langle X_{\max} \rangle$ respectively.}, and the all-particle spectrum $J(E)$~\cite{PierreAuger:2021hun}. Data points from observables $J(E)$, $\langle X_{\max} \rangle$, and $\sigma\left( X_{\max} \right)$ are included in the likelihood function at energies $E > 10^{18.7}$~eV, while the H spectrum is included in the likelihood function at energies $10^{17.8}$~eV~$\leq E \leq 10^{18.7}$~eV.

The difference between the data $d_k$ with statistical uncertainty $\sigma_k$ (if the statistical uncertainty is asymmetrical in bin $k$, we take the upper uncertainty if the model lies above the data point and the lower uncertainty otherwise) in energy bin $E_k$ with the model-predicted value of $m_k$ is assumed to be normally distributed, so the natural logarithm of the modified likelihood function\footnote{In our definition of the modified likelihood function, given that \texttt{emcee} is not sensitive to the normalisation, we omit the constant Gaussian prefactor $(2 \pi \sigma_{k}^{2})^{-1/2}$.} across all data points is defined as
    \begin{equation}
        \ln L_m = -\frac{1}{2} \sum\limits_{k} \left(\frac{d_k - m_k}{\sigma_k}\right)^{2},
    \end{equation}
where the sum runs across all energy bins of $E^{3} J(E)$, $\langle X_{\max} \rangle$, $\sigma\left( X_{\max} \right)$, and the H spectrum. In Ref.~\cite{PierreAuger:2025eun}, the Pierre Auger Collaboration reports two upper limits of $J(E)$ in two energy bins, which we also take into account, assuming the Poisson distribution for zero observed events under some model-expected value:
    \begin{equation}
        \ln L_{\mathrm{UL}} = -n_{\mathrm{exp, \, UL \,}1} - n_{\mathrm{exp, \, UL \,}2},
    \end{equation}
where $n_{\mathrm{exp, \, UL \,}k}$ is the model-predicted number of events in the last or second-to-last energy bin for the constant exposure of $1.05 \times 10^{5}$ km$^{2}$ sr yr \cite{PierreAuger:2025eun}. The total logarithm of likelihood is the sum of the logarithm of the ``likelihood of measurements'' and the logarithm of the ``likelihood of upper limits''.

We run the MCMC long enough so that it converges to a state around the maximum of the likelihood function. To ensure convergence, we calculate the Gelman-Rubin diagnostic $\hat{R}$. If $\hat{R} < 1.05$, the MCMC is thought to have converged: this is verified a posteriori for each fit. As the best-fit values in our plots, we use those parameter values providing the maximum a posteriori value of the posterior function $L_{\mathrm{MAP}}$, and, since the prior distribution is uninformative, the values of the posterior distribution of accepted MCMC steps are equal to the values of the likelihood function at its maximum: $L_{\mathrm{MAP}} = L_{\mathrm{ML}}$.

To assess the goodness of fit, we calculate deviance $-2 \ln L_{\mathrm{MAP}} / \mathrm{ndf}$, which is asymptotically $\chi^{2}$-distributed. Here, $\mathrm{ndf}$ is the number of degrees of freedom equal to the number of measured data points plus two (for the spectral upper limits) minus the number of model parameters (eight).

\section{Elemental Abundances of Galactic Cosmic Rays}\label{sec:gcr_fractions}
The fractions for GCRs are taken from Table 1 of Ref.~\cite{Hoerandel:2002yg} and assuming a given universal index of GCRs of $\gamma_{\mathrm{GCR}} = 2.6$ when converting them to the fractions at a \textit{fixed rigidity} with Eq.\,(\ref{eq:fractions_energy_to_rigidity}). We sum up fluxes of C, N, O to calculate the effective Galactic fraction of N; fluxes of elements between F ($Z = 9$) and Si ($Z = 14$) are summed up to calculate the effective Galactic fraction of Si; the sum of fluxes of elements between P ($Z = 15$) and Fe ($Z = 26$) is used for the effective Galactic fraction of Fe.

The spectrum of different nuclei of GCRs in the energy range between $10^{11}$ eV and $10^{14}$ eV does not obey a simple power law with a universal spectral index \cite{DAMPE:2025opn}. However, utilising reference fluxes at 1 TeV from Table 1 in \cite{Hoerandel:2002yg} and assuming $\gamma_{\mathrm{GCR}} = 2.6$, we find that a typical relative error with respect to the data obtained by the DAMPE Collaboration \cite{DAMPE:2025opn} is around $0.2$. Namely, the medians of the absolute values of the relative errors are $0.07$ for hydrogen, $0.21$ for helium, $0.19$ for nitrogen, and $0.20$ for iron. Similar values are obtained for the difference between phenomenological models \texttt{polygonato} \cite{Hoerandel:2002yg}, \texttt{GST} \cite{Gaisser:2013bla}, \texttt{GSF-2025} \cite{Dembinski:2025nmp,Fedynitch:2026ugq} and the fluxes measured in Ref.~\cite{DAMPE:2025opn}.

Another source of uncertainty in elemental fractions of GCRs is the bias due to the observation of low-energy CRs (in our case, GCRs) only from one galaxy---the Milky Way. Other galaxies have very similar abundances of hydrogen and helium, but somewhat varying abundances of heavier elements, with fluctuations of around $0.1$ dex in massive galaxies (see, e.g., Fig. 15 in Ref.~\cite{MaiolinoandMannucci2019}).

Thus, combining these two uncertainties quadratically, we adopt an approximate relative uncertainty of GCR fractions as $\pm 30$\%.

\section{Systematic Uncertainties}\label{sec:systematics}
\begin{figure}[t]
        \includegraphics[width=\linewidth]{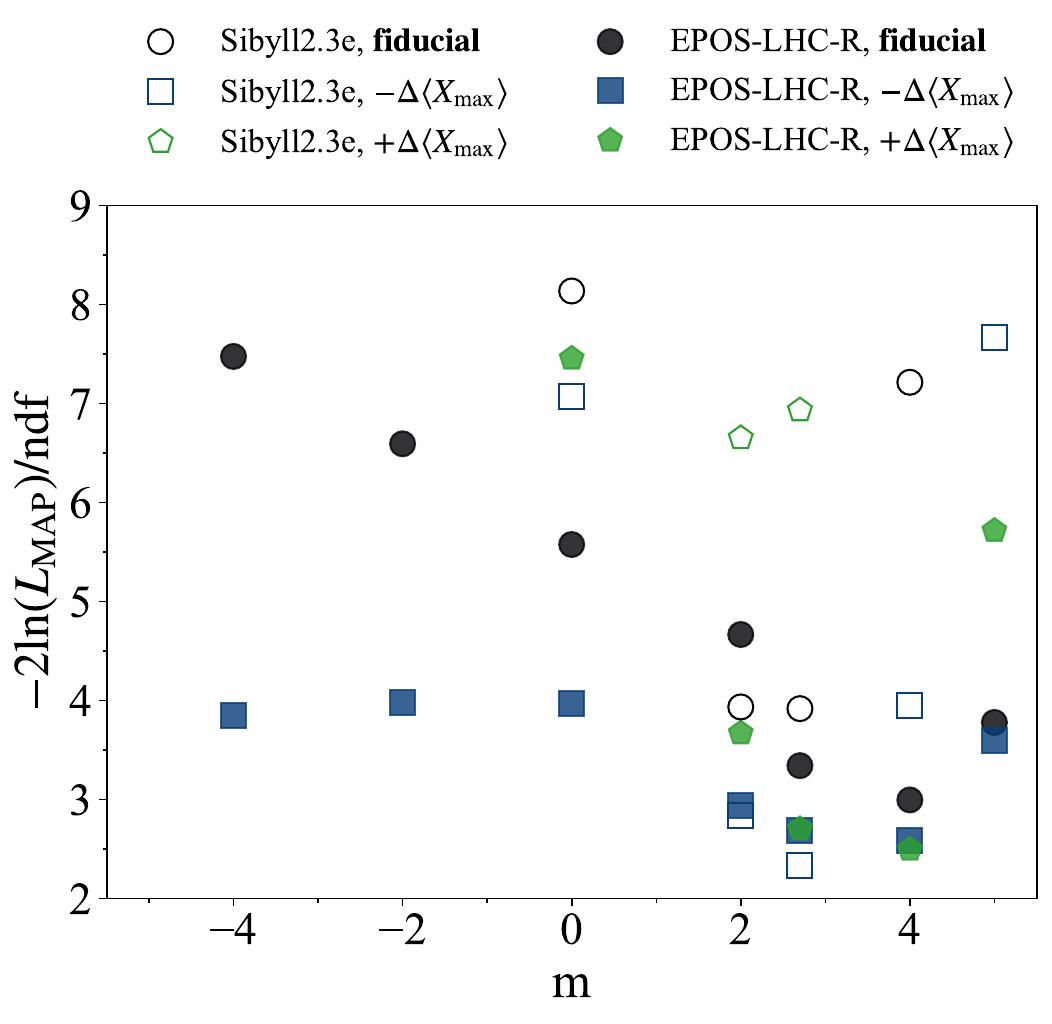}
        \caption{
        \justifying
            Goodness of fits with various values of the exponent $m$ of the cosmological evolution of UHECR sources for \texttt{EPOS-LHC-R} and \texttt{Sibyll 2.3e} and three sets of \textit{Auger} data (one fiducial dataset with no shifts, and two datasets with shifts $\pm \Delta \langle X_{\max} \rangle$ by one systematic uncertainty). Fits with the goodness $-2 \ln L_{\mathrm{MAP}} / \mathrm{ndf}$ worse than $9$ are not shown.
        }
        \label{fig:goodness}
    \end{figure}
    \begin{figure}[t]
        \centering
        \includegraphics[width=\linewidth]{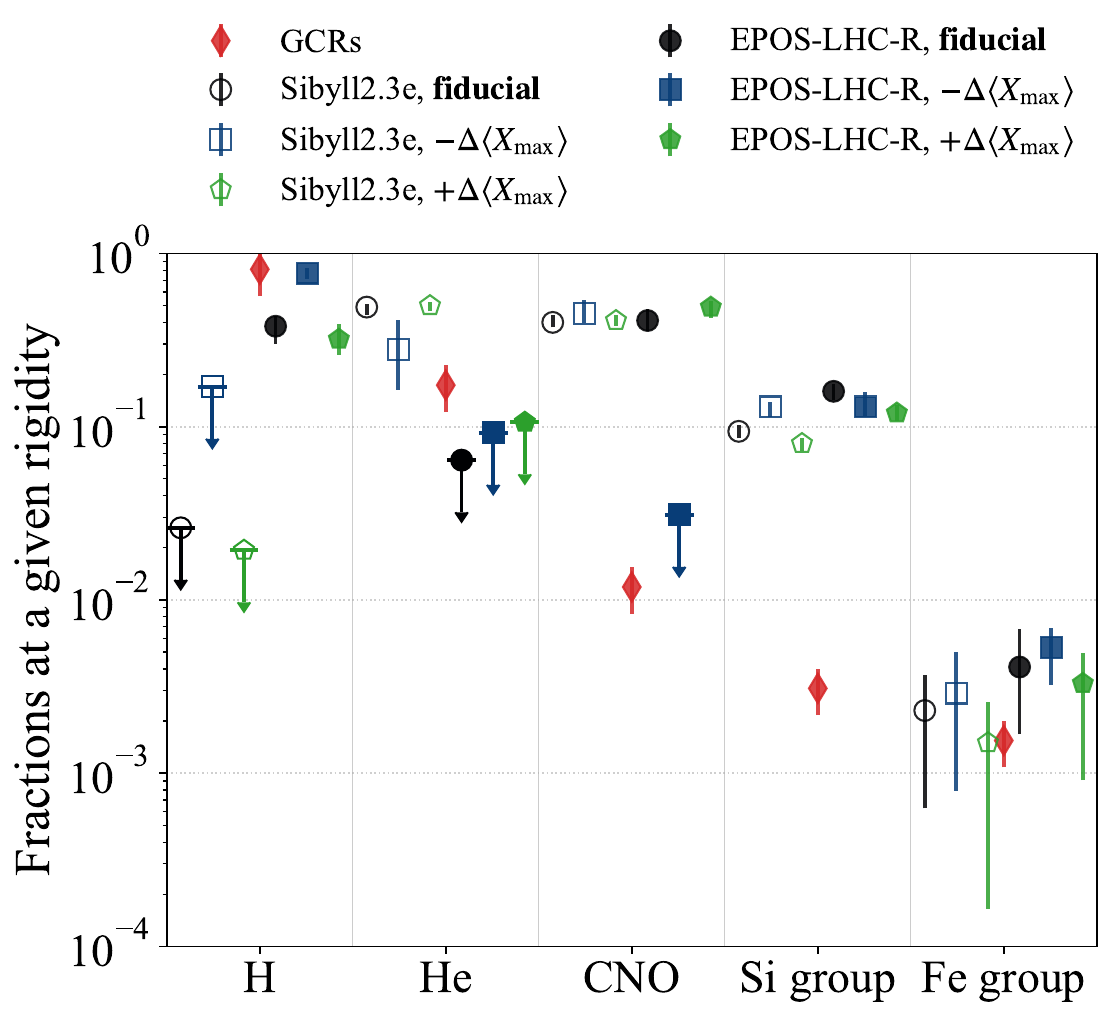}
        \caption{
        \justifying
            Credible intervals (68\%, the highest-density) for fractions $\rho$ at a given \textit{rigidity} obtained with \texttt{EPOS-LHC-R} and \texttt{Sibyll 2.3e} for one fiducial dataset with no shifts, and two datasets with shifts $\pm \Delta \langle X_{\max} \rangle$ up/down by one systematic uncertainty. UHECR sources are assumed to follow the star formation rate.
        }
        \label{fig:fractions_at_a_fixed_rigidity_epos_lhc_r_and_sibyll_2_3_e}
    \end{figure}
The systematic uncertainty of the $\langle X_{\max}\rangle$ \textit{Auger} measurements exceeds the statistical one, whereas statistical errors of $\sigma(X_{\max})$ are larger than the systematic ones~\cite{PierreAuger:2026qbt}. We investigate the effect of the systematic uncertainty in $\langle X_{\max}\rangle$ by performing two additional fits (for each value of power $m$ of the UHECR source evolution function) with data points of $\langle X_{\max}\rangle$ shifted by $\pm1\,\sigma_{\mathrm{syst}}$ (individually in each energy bin).

To assess the effects of uncertainties in modelling the development of EASs with different HIMs, we performed fits with the same three perturbations of the \textit{Auger} data using \texttt{Sibyll\,2.3e} as a test alternative model\footnote{We use the \texttt{exp} parametrisation of this HIM from Ref.~\cite{Evoli:2026bea}, see their Table B1.}.

The goodness of all performed fits is shown in Fig.~\ref{fig:goodness} as a function of power $m$ of the UHECR source evolution function
    \begin{equation}
        f(z,m) \propto \begin{cases}
            (1 + z)^{m}, \mathrm{\,if \,} m \leq 0, \, z \leq z_{\max},\\
            (1 + z)^{m}, \mathrm{\,if \,} m > 0, \, z \leq z_{0},\\
            (1 + z_{0})^{m}, \mathrm{\,if \,} m > 0, \, z_{0} < z \leq z_{\max},\\
            0, \mathrm{\,if \,} z > z_{\max},
        \end{cases}
    \label{eq:cosmolution}
    \end{equation}
where $z_{\max} = 4$, $z_{0} = 1.5$, and for $m = 2.7$, which is close to the cosmological evolution of sources following the star formation rate, we use Eq.~(15) of Ref.~\cite{Madau:2014bja}.

\begin{table}[t]
        \caption{\label{table:credible_intervals_CENTRE_GROUND}
            \justifying
            Comparison of the modes of the marginal posterior distributions and the highest-density $68$\% credible intervals obtained with \texttt{EPOS-LHC-R} and \texttt{Sibyll\,2.3e} for the nominal \textit{Auger} data. Parameter $\rho_{\mathrm{H}}$ is derived from the four other fractions.
        }
        \begin{ruledtabular}
        \begin{tabular}{r|r|r}
            Parameter & \texttt{EPOS-LHC-R} & \texttt{Sibyll\,2.3e}\\
            \hline
            $\gamma$ & $0.81^{+0.06}_{-0.05}$ & $-0.72^{+0.20}_{-0.20}$ \\
            $R_{\mathrm{cut}}$ [$10^{18}$ V] & $8.1^{+0.3}_{-0.5}$ & $3.4^{+0.5}_{-0.5}$ \\
            $\lambda$ & $2.9^{+0.3}_{-0.2}$ & $1.40^{+0.14}_{-0.09}$ \\
            $\rho_{\mathrm{H}}$ & $0.38^{+0.05}_{-0.08}$ & $8^{+253}_{-8} \times 10^{-4}$ \\
            $\rho_{\mathrm{He}}$ & $0.00^{+0.06}_{-0.00}$ & $0.49^{+0.02}_{-0.05}$ \\
            $\rho_{\mathrm{N}}$ & $0.41^{+0.07}_{-0.05}$ & $0.40^{+0.04}_{-0.03}$ \\
            $\rho_{\mathrm{Si}}$ & $0.160^{+0.018}_{-0.017}$ & $0.094^{+0.008}_{-0.008}$ \\
            $\rho_{\mathrm{Fe}}$ & $0.0041^{+0.0027}_{-0.0024}$ & $0.0023^{+0.0014}_{-0.0017}$ \\
            \hline
            $-2\ln L_{\mathrm{MAP}} / \mathrm{ndf}$ & 3.34 & 3.92 \\
        \end{tabular}
        \end{ruledtabular}
    \end{table}
    \begin{table}[t]
        \caption{\label{table:credible_intervals_CENTRE_DOWN}
            \justifying
            Same as Table~\ref{table:credible_intervals_CENTRE_GROUND}, but for the fits with $\Delta X_{\mathrm{max}} = -1 \sigma_{\mathrm{syst}}$.
        }
        \begin{ruledtabular}
        \begin{tabular}{r|r|r}
            Parameter & \texttt{EPOS-LHC-R} & \texttt{Sibyll\,2.3e}\\
            \hline
            $\gamma$ & $1.00^{+0.08}_{-0.03}$ & $0.34^{+0.19}_{-0.11}$ \\
            $R_{\mathrm{cut}}$ [$10^{18}$~V] & $6.0^{+0.6}_{-0.7}$ & $6.2^{+0.8}_{-0.7}$ \\
            $\lambda$ & $1.7^{+0.4}_{-0.1}$ & $2.0^{+0.4}_{-0.2}$ \\
            $\rho_{\mathrm{H}}$ & $0.77^{+0.05}_{-0.06}$ & $0.05^{+0.12}_{-0.05}$ \\
            $\rho_{\mathrm{He}}$ & $0.02^{+0.08}_{-0.02}$ & $0.28^{+0.14}_{-0.12}$ \\
            $\rho_{\mathrm{N}}$ & $0.003^{+0.028}_{-0.003}$ & $0.45^{+0.09}_{-0.05}$ \\
            $\rho_{\mathrm{Si}}$ & $0.130^{+0.029}_{-0.015}$ & $0.130^{+0.009}_{-0.017}$ \\
            $\rho_{\mathrm{Fe}}$ & $0.0053^{+0.0016}_{-0.0021}$ & $0.0029^{+0.0021}_{-0.0021}$ \\
            \hline
            $-2\ln L_{\mathrm{MAP}} / \mathrm{ndf}$ & 2.69 & 2.34 \\
        \end{tabular}
        \end{ruledtabular}
\end{table}
    
None of the fits with $m \leq 0$ is satisfactory, since, due to a dominant contribution of nearby sources, not enough secondary nucleons are produced during the intergalactic propagation of heavy primary nuclei, and primary H can not alone fit the shape of the H spectrum observed by \textit{Auger}. Except for the two fits with $m = 2$ and for the star formation rate ($m = 2.7$) with $\Delta \langle X_{\max}\rangle = -1 \sigma_{\mathrm{syst}}$, \texttt{EPOS-LHC-R} fits the data better than \texttt{Sibyll\,2.3e}. Fits with \texttt{Sibyll\,2.3e} result in a higher inferred metallicity of UHECRs. Moreover, we find two solutions (with dominant secondary protons and with comparable contributions of secondary and primary H nuclei below $10^{18.7}$~eV) only with \texttt{EPOS-LHC-R}. Since the latter solution results in a more conservative estimate of the UHECR metallicity, in the main text, we present the results obtained with \texttt{EPOS-LHC-R}.

We plot credible intervals for fractions of UHECR nuclei at a given \textit{rigidity} inferred with \texttt{Sibyll\,2.3e} and \texttt{EPOS-LHC-R} for the three perturbations of the \textit{Auger} data in Fig.~\ref{fig:fractions_at_a_fixed_rigidity_epos_lhc_r_and_sibyll_2_3_e} (for the UHECR source evolution following the star formation rate). We conclude that the fractions of silicon nuclei in UHECRs at a given \textit{rigidity} exceed those of GCRs by more than an order of magnitude for both \texttt{EPOS-LHC-R} and \texttt{Sibyll\,2.3e} HIMs, and when taking into account the systematic uncertainties of the \textit{Auger} energy scale and $\langle X_{\max}\rangle$ measurements.

Credible intervals for eight parameters inferred with \texttt{EPOS-LHC-R} and \texttt{Sibyll\,2.3e} for the nominal \textit{Auger} data are compared in Table~\ref{table:credible_intervals_CENTRE_GROUND}, while in Table~\ref{table:credible_intervals_CENTRE_DOWN} we present a comparison for the fits with the measured $\langle X_{\max}\rangle$ shifted down by $\Delta \langle X_{\max} \rangle = -1 \sigma_{\mathrm{syst}}$. We note that a superexponential cutoff is always significantly preferred over the classical exponential cutoff, and that fits with \texttt{Sibyll\,2.3e} require significantly harder spectra than those with \texttt{EPOS-LHC-R}.

\section{Effect of the Source Evolution on the Observable UHECR Spectra}
\label{sec:source_evolution_effect}
    \begin{figure}[t]
        \centering
        \includegraphics[width=\linewidth]{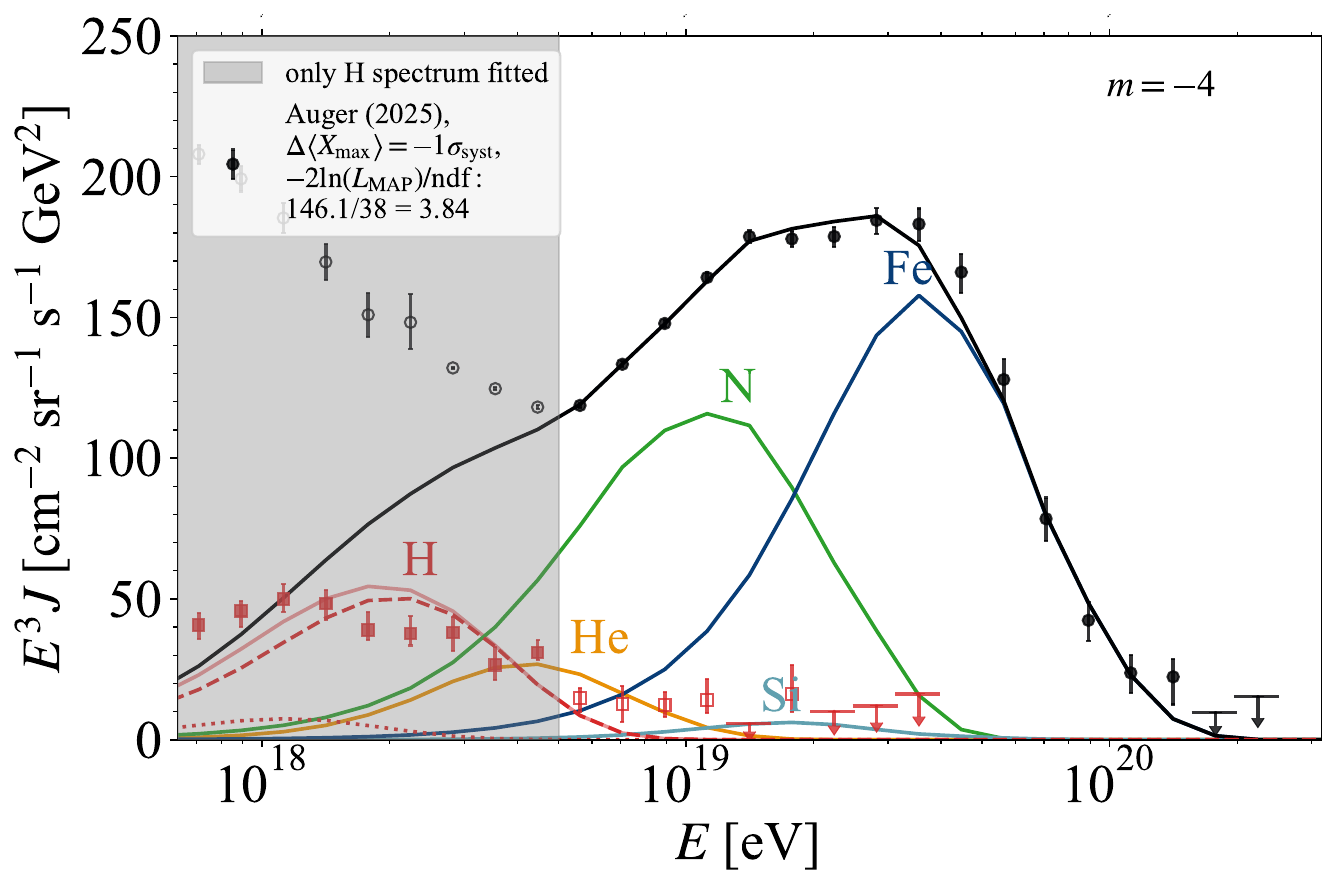}
        \includegraphics[width=\linewidth]{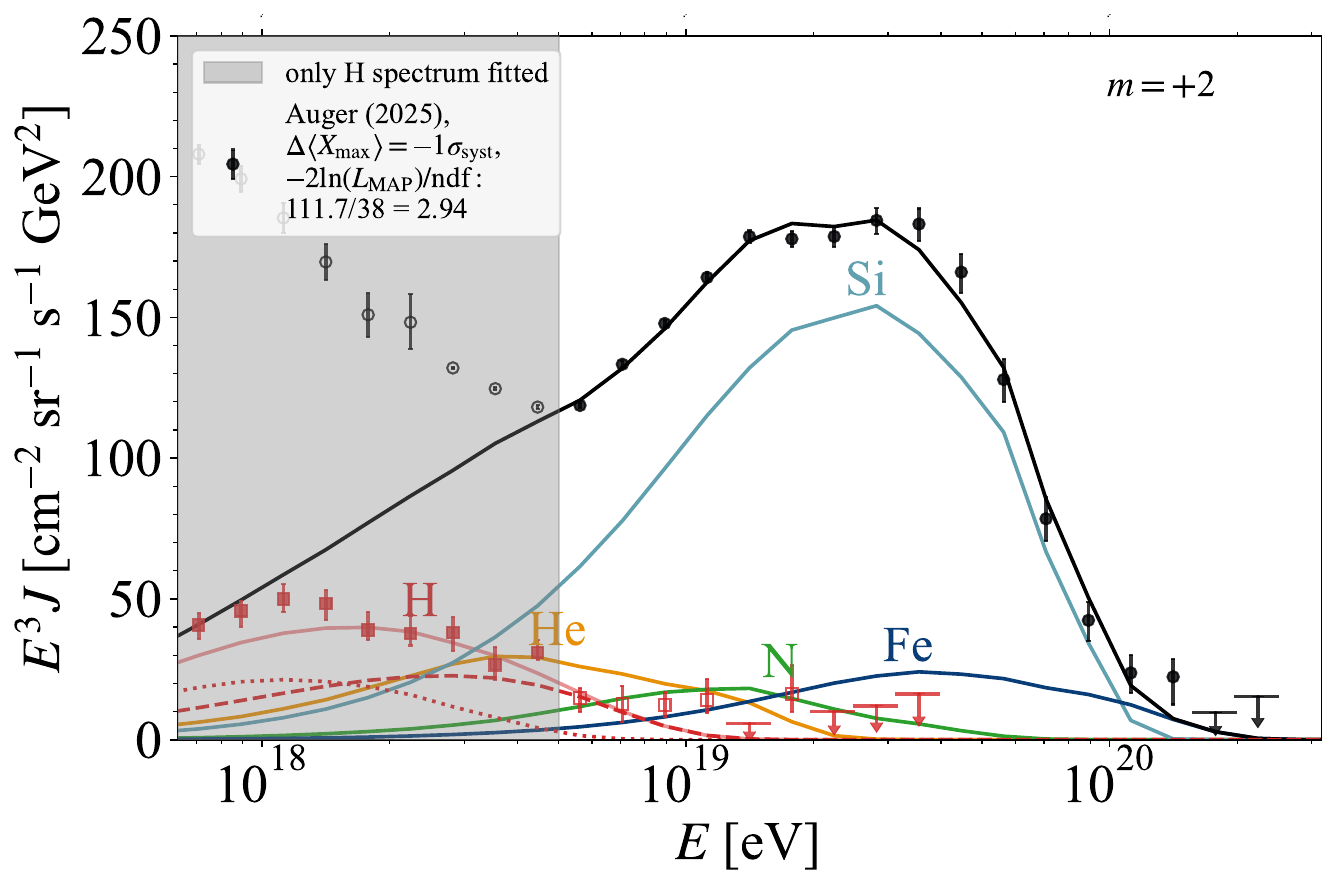}
        \includegraphics[width=\linewidth]{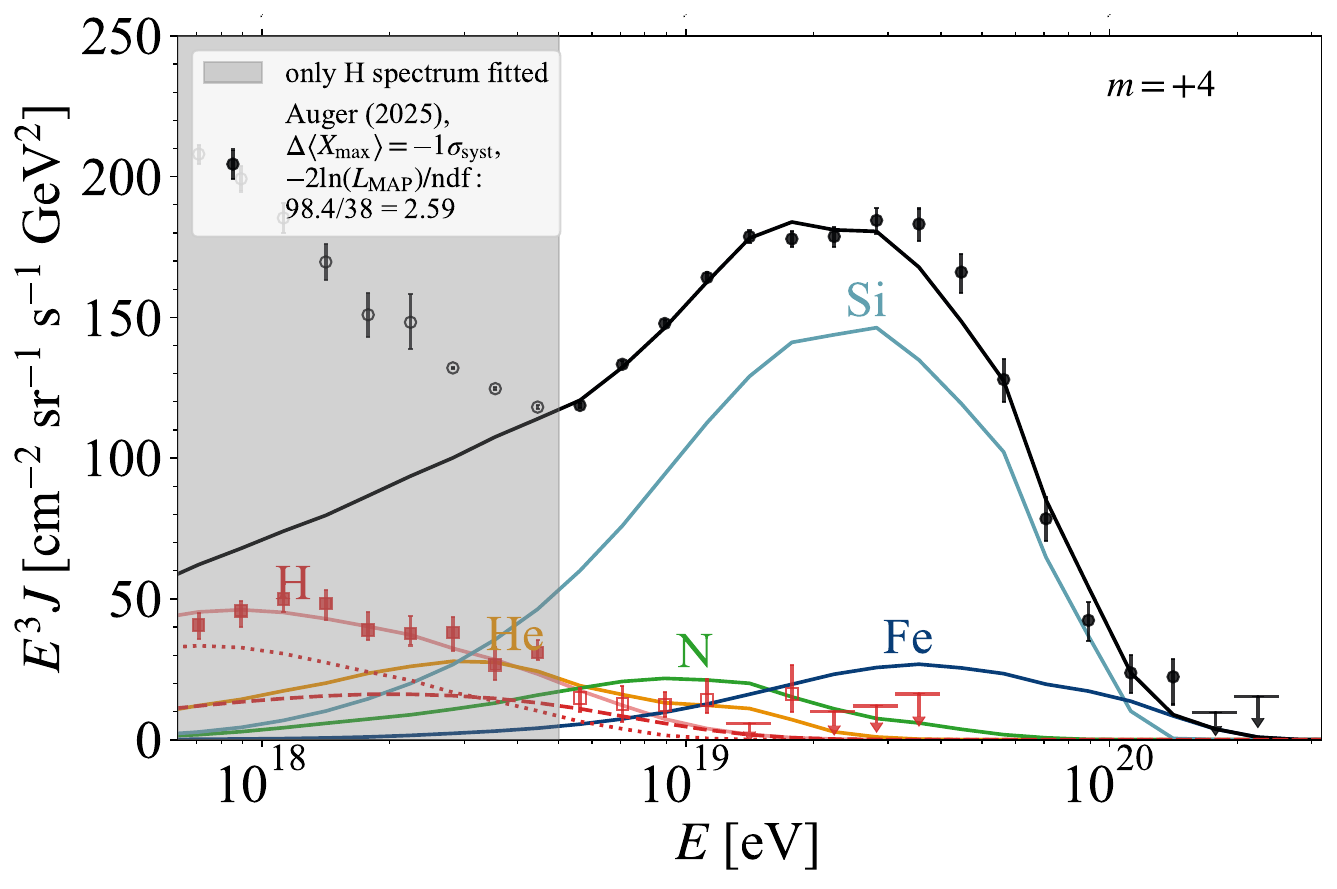}
        \caption{
            \justifying
            Intensity times the cube of energy of observed and simulated cosmic rays in linear-log scale. Secondary protons are shown as dotted lines while primary protons are shown as dashed lines. $\Delta \langle X_{\max} \rangle = -1 \sigma_{\mathrm{syst}}$. Top panel: $m=-4$; middle panel: $m=+2$; lower panel: $m=+4$.
        }
        \label{fig:fitting_with_and_without_H_fractions}
    \end{figure}
    Fig.~\ref{fig:fitting_with_and_without_H_fractions} illustrates the effect of the source evolution on the all-particle and H spectra observable at Earth. The negative source evolution ($m = -4$) makes the contribution of nearby sources dominant, so that heavy nuclei escaping from them reach Earth almost without interacting with the EBL, hence not producing any significant amount of secondary nucleons. The H spectrum observed by \textit{Auger} seems to be too broad to be described only by primary H nuclei, as can be seen in the upper panel of Fig.~\ref{fig:fitting_with_and_without_H_fractions}. The observed lack of H nuclei at low energies leads to poor quality of the fits obtained for $m \leq 0$, as illustrated in Fig.~\ref{fig:goodness}. A more positive source evolution with $m = +2$ allows a better fit to the data with comparable contributions of secondary and primary protons peaking (in the $E^{3}J$ plot) at different energies (the middle panel of Fig.~\ref{fig:fitting_with_and_without_H_fractions}). Even more positive evolution with $m = +4$ results in a slightly better fit, with secondary H nuclei saturating the observed \textit{Auger} H spectrum at $E \lesssim 10^{18}$~eV (the lower panel of Fig.~\ref{fig:fitting_with_and_without_H_fractions}). When $m \geq 5$, the secondary nucleons due to photodisintegration of heavier primary nuclei start overshooting the observed \textit{Auger} H spectrum, resulting in the worsening of the fits seen in Fig.~\ref{fig:goodness}.

\section{Corner Plots}
\begin{figure*}[t]
    \centering
    \includegraphics[width=\linewidth]{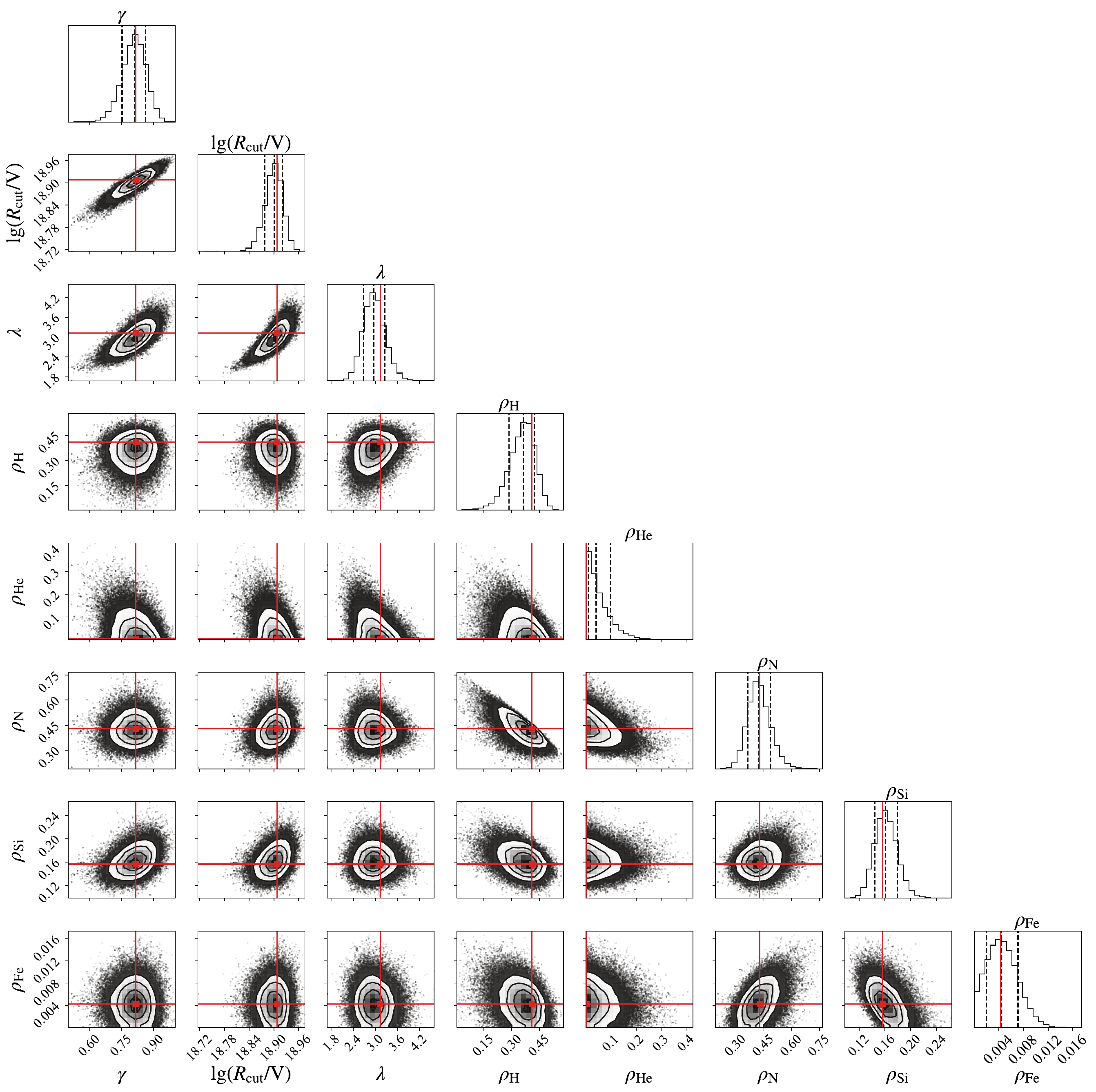}
    \caption{
    \justifying
        Corner plot with the posterior model parameter distributions for our model with \texttt{EPOS-LHC-R}, no data shifts and UHECR sources following the star-formation rate. The hydrogen fraction $\rho_{\mathrm{H}}$ at a given \textit{rigidity} is a derived parameter (as unity minus the sum of the four other fractions). Colourful horizontal and vertical lines indicate the values providing the maximum value of the a posteriori probability function. Dashed vertical lines in the 1D marginal histograms represent 16\%, 50\%, and 84\% percentiles of the posterior marginal distributions. Contour lines in the 2D marginal histograms represent 11.8\%, 39.3\%, 67.5\%, and 86.4\% 2D containment regions. Black dots represent individual walkers outside the last containment region.
    }
    \label{fig:corner_EPOS-LHC-R_CENTRE_GROUND}
\end{figure*}

\begin{figure*}[t]
    \centering
    \includegraphics[width=\linewidth]{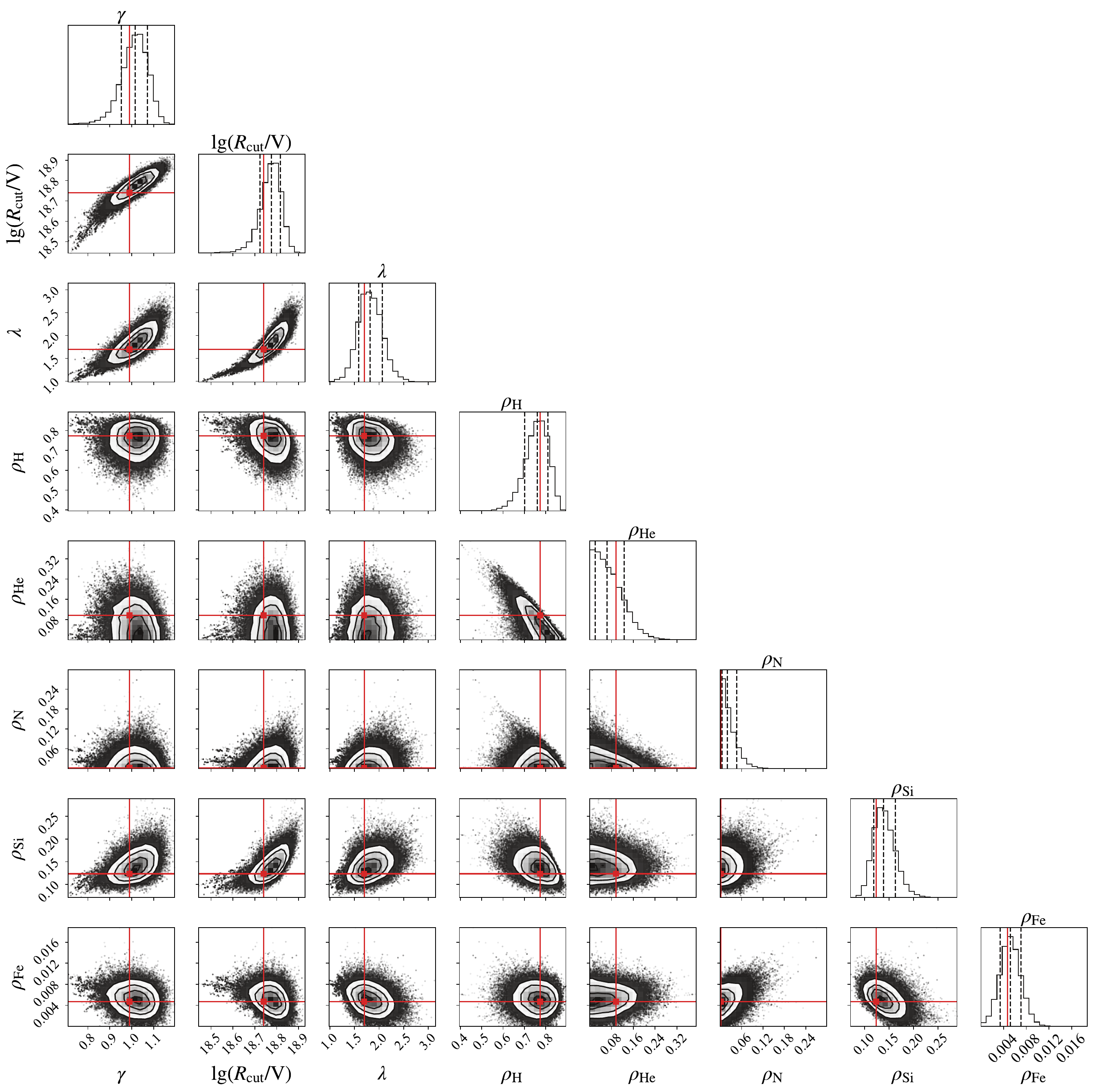}
    \caption{
    \justifying
        Same as Fig.\ \ref{fig:corner_EPOS-LHC-R_CENTRE_GROUND}, but for the fit with the \textit{Auger} data shift $\Delta \langle X_{\max} \rangle = -1 \sigma_{\mathrm{syst}}$, which results in the primary-H-enriched scenario and a lower value of the inferred UHECR metallicity $Z_{\rho}$.
    }
    \label{fig:corner_EPOS-LHC-R_CENTRE_DOWN}
\end{figure*}

    We show here corner plots with marginalised posterior distributions of the model parameters for the two solutions (for the UHECR source evolution following the star formation rate), discussed in the main text, in Figs.~\ref{fig:corner_EPOS-LHC-R_CENTRE_GROUND} and \ref{fig:corner_EPOS-LHC-R_CENTRE_DOWN}.

\end{document}